\documentclass[10pt,journal,epsfig]{IEEEtran}

\usepackage[dvips]{graphicx}
\usepackage{graphicx}
\usepackage{amssymb}
\usepackage{cite}
\usepackage{subfigure}
\usepackage{amsmath}

\usepackage{subeqnarray}
\usepackage{cases}
\usepackage{color}
\usepackage[centerlast]{caption2}
\usepackage{wrapfig}
\usepackage{url}

\usepackage{siunitx} % Required for alignment
\usepackage{multirow} % Required for multirows
\usepackage{booktabs} % For prettier tables
\usepackage{longtable} % To display tables on several pages
\usepackage{rotating} % To display tables in landscape

\usepackage{algorithm}
\usepackage[noend]{algpseudocode}

\begin{document}

\IEEEoverridecommandlockouts
\title{An Incentive-Aware Job Offloading Control Framework for Mobile Edge Computing
}
\author{ Lingxiang Li ~\IEEEmembership{Member,~IEEE}, Tony Q.S. Quek ~\IEEEmembership{Fellow,~IEEE}, Ju Ren ~\IEEEmembership{Member,~IEEE} \\
Howard H. Yang ~\IEEEmembership{Member,~IEEE}, Zhi Chen ~\IEEEmembership{Senior Member,~IEEE}, Yaoxue Zhang ~\IEEEmembership{Senior Member,~IEEE}
\thanks{Lingxiang Li, Ju Ren and Yaoxue Zhang are with School of Information Science and Engineering, Central South University, Changsha 410083, China
(e-mails: \{lingxiang.li, renju, zyx\}@csu.edu.cn).}
\thanks{Tony Q.S. Quek and Howard H. Yang are with Information Systems Technology and Design Pillar, Singapore University of Technology and Design, Singapore 487372
(e-mails: \{tonyquek, howard\_yang\}@sutd.edu.sg).}
\thanks{Zhi Chen is with the National Key Laboratory of Science and Technology on Communications,
UESTC, Chengdu 611731, China (e-mail: chenzhi@uestc.edu.cn).}
}

\maketitle

\begin{abstract}
This paper considers a scenario in which an access point (AP) is equipped with a mobile edge server of finite computing power,
and serves multiple resource-hungry mobile users by charging
users a price. Pricing provides users with incentives in offloading.
However, existing works on pricing are based on abstract concave utility functions (e.g, the logarithm function),
giving no dependence on physical layer parameters.
To that end, we first introduce a novel utility function, which measures
the cost reduction by offloading as compared with executing jobs locally.
Based on this utility function we then formulate two offloading games,
with one maximizing individual's interest and the other maximizing the overall system's interest.
We analyze the structural property of the games and admit in closed form
the Nash Equilibrium and the Social Equilibrium, respectively.
The proposed expressions are functions of
the user parameters such as the weights of
computational time and energy, the distance from the AP,
thus constituting an advancement over prior economic works that have considered only abstract functions.
Finally, we propose an optimal pricing-based scheme, with which
we prove that
the interactive decision-making process with self-interested users
converges to a Nash Equilibrium point equal to the Social Equilibrium point.
\end{abstract}


\section{Introduction}
With the development of internet services, a diverse variety of computing-intensive applications such as
mobile shopping, face recognition, and augmented reality, are emerging and attracting great attention.
The execution of those novel applications typically requires low latency and high power consumption \cite{SLiu13}.
Meanwhile, due to its physical size constraint, the mobile device usually has insufficient
local computing power or limited battery capacities.
As such, it may not provide satisfactory quality of experience to the users.
An alternative way is to offload all or part of the jobs to the resource-rich cloud infrastructures such as Amazon EC2
via the access point (AP) associated with users.
Such cloud-aided computing eases users' burden in executing computing-intensive jobs in time. However, it also suffers
from wide area network (WAN) delay between the cloud and the AP, as the cloud is often located far
away from the users \cite{MSat09}.

More recently, a new trend of moving the function of cloud computing to the network edge
is happening. Along this line and in the cellular network, the network edge refers to the radio access network (RAN) side of the Internet,
and the AP (e.g., the base station, the Wi-Fi) is equipped with one additional
computing server, i.e., the mobile edge computing (MEC) server.
There are two dominant advantages of applying MEC technology in cellular networks.
First, instead of residing in a far-away cloud, the computing server is located at the AP, close to
the mobile device users, thus avoiding the WAN delay between the AP and the cloud.
Second, since the AP is in control of both the computing and the radio resource, a joint optimization
of those resources could be performed, bringing about considerable
improvement in the computing/radio resource efficiency.
 
Presently, researchers have been actively studying the joint optimization of computing and radio resources for
different MEC based scenarios. Specifically, the works of \cite{Yuyi16,Changsheng16,Juan16,Yunzheng18} consider the single user
case, jointly deciding the offloading decision, the CPU-cycle frequencies for mobile execution, and the transmit power for offloading.
The works of \cite{Yuyi166,Yuyi17,Changsheng17,Chen17} further examine the multi-user case, with users offloading jobs
to the AP in a spatial division multiple access (SDMA), a time division multiple access (TMDA) or a frequency division multiple access (FDMA) mode. The case with multiple access points is studied in \cite{Thinh17}, wherein
a mobile user allocates its jobs to multiple nearby access points.
The D2D case is studied in \cite{Lingjun16}, wherein mobile users can dynamically and beneficially share the computation and
communication resources among each other.
More complex energy harvesting driven networks are considered in \cite{Fei18,Fengxian18},
where energy-constrained mobiles are considered and the energy harvesting technique is integrated into the mobile edge computing cellular networks.
The above studies all formulate a centralized resource allocation problem, which involves solving a Mixed Integer Nonlinear Programming (MINP) problem,
and most of them only give numerical results to illustrate the performance.

Due to the nature of sharing an MEC server, the expected delays of different users
spent in the system required by the edge computing are coupled.
Hence, it is usually very complicated to optimize the MINP problem centrally.
Moreover, since users are strategic, making decisions to maximize
individual welfare, it is not easy to control the decision variable directly.
Therefore, another line of research targets resource allocation in a distributed way.
In particular, the works of \cite{Xuchen15,Xuchen16,Guo16,Jianchao18,Ling16}
reformulate the original centralized MINP problem into an interactive decision-making process among a group of rational users.
Based on game theory the equilibrium points of the interactive decision
process are examined, and decentralized algorithms with low complexity are proposed.
In \cite{Lyu17}, the original problem is decomposed into several
easy-to-handle sub-problems and solved semi-distributively with convex optimization techniques.

\subsection{Related works and main contributions of this paper}
In this paper, we aim to do resource allocation in a decentralized way.
Different from \cite{Xuchen15,Xuchen16,Guo16,Jianchao18,Ling16,Lyu17}, we take queueing delays into account.
Specifically, we consider an MEC assisted RAN serving multiple mobile device users as in \cite{Lyu17}, except that, unlike \cite{Lyu17}
which only considers a binary decision of offloading, we consider a more general case of controlling the flow rate of offloading.
Our network comprises users that work in a FDMA mode to offload part/all of its jobs to the AP. Users may have different
attributes, such as distance from the AP, battery capacity, and the application it runs.
As such, different users have different utilities of offloading. In addition, since the computing power of the edge is limited,
the users offloading jobs are competing for computing resource, and congestion may happen if too many jobs are offloaded to the edge. Therefore,
waste would be reduced if the limited computing resources at the AP are reserved for the users that value them the most.

Pricing is a key tool to allocate resources to the users who value them the most, thus controlling congestion in resource competition cases \cite{Courcoubetis03,Huang13}.
In addition, in economics there are some mechanisms such as the Vickrey-Clarke-Groves (VCG) mechanism,
which is designed to provide incentives to the users to choose the socially optimal levels of demand and reveal their true
utility functions; this mechanism is especially useful when the allocation requires users' payoff functions and users may have motivations to cheat.
However, existing works from the economics viewpoint such as \cite{Weiwei18,Long16} are based on abstract utility functions (e.g., a simple logarithm function of the amount of product) and
give no specific dependence of the utility function on the physical layer parameters. In this paper,
we study the interactive decision-making process among users to optimize the physical layer parameters.
Unlike the works of \cite{Xuchen15,Xuchen16,Guo16,Jianchao18,Ling16}, which only study the Nash Equilibrium
and no incentive-aware scheme is considered, in this paper we aim to integrate the ideas of economics into the optimization of
MEC networks. Our goal is to provide users with the appropriate incentives to control their flows
and improve the overall system performance in a distributed way.
Our main contributions are summarized below.
\begin{enumerate}
\item
We introduce a novel utility function (see eq. (\ref{eq13})), which subtracting the delay cost
required by edge computing equals the profit a user achieves by offloading (\ref{eq12b}).
This utility function measures %the cost reduction by offloading as compared with executing jobs locally, i.e.,
the difference between the local computing cost and the offloading cost (including the offloading delay and the offloading energy consumption).
We prove that the proposed utility function is concave and monotonously increasing with respect to users' offloading flow rate,
satisfying the basic nature of the utility function in economics, thus
bridging the gap between the economics and physical layer parameter optimizations.
\item Based on the given utility function we propose an incentive-aware job offloading framework, in which
the user decides the amount to offload by comparing its utility function with the delay cost required by edge computing.
We determine in closed form the Nash Equilibrium point (see \emph{Theorem} \ref{theorem1})
as well as the Social Equilibrium point (see \emph{Theorem} \ref{theorem2}).
Our analytical results fully describe the dependence of the Equilibria on the edge computing power as well as
the user parameters such as the local computing power, the weights of the
computational time and energy the distance from the AP, and apply to any number of users.
\item We derive in closed form the sufficient and necessary conditions for a positive unique
Nash/Social Equilibrium to exist (see \textit{Corollary} \ref{corollary1} and \textit{Corollary} \ref{corollary2}, respectively).
We also propose an optimal pricing-based scheme (see {Table Algorithm} \ref{alg:NE}),
which regulates users' action by adding an appropriate price.
With this pricing-based scheme, the individual objectives of self-interested users are aligned with that of the overall system, and
the interactive decision-making process with self-interested users
converges to a Nash Equilibrium point equal to the Social Equilibrium point (see \emph{Theorem} \ref{theorem3}).
\end{enumerate}

\subsection{Organization of this paper}
The rest of this paper is organized as follows.
In Section II, we introduce the system model, including the computation model, the
radio access model and the offloading policy. In Section III, we propose a business model for
mobile edge computing, based on which we formulate two games, with one maximizing individual's
interest and the other maximizing the overall system interest. In Section IV, we
examine the equilibrium points of the games, based on which we propose an optimal pricing-based scheme.
We prove that with this pricing-based scheme, the individual objectives
of self-interested users are aligned with that of the overall system.
Numerical results are given in Section V and conclusions are drawn in Section VI.

\newtheorem{proposition}{Proposition}
\newtheorem{theorem}{Theorem}
\newtheorem{corollary}{Corollary}
\newtheorem{lemma}{Lemma}

\begin{figure}[!t]
\centering
\includegraphics[width=3in]{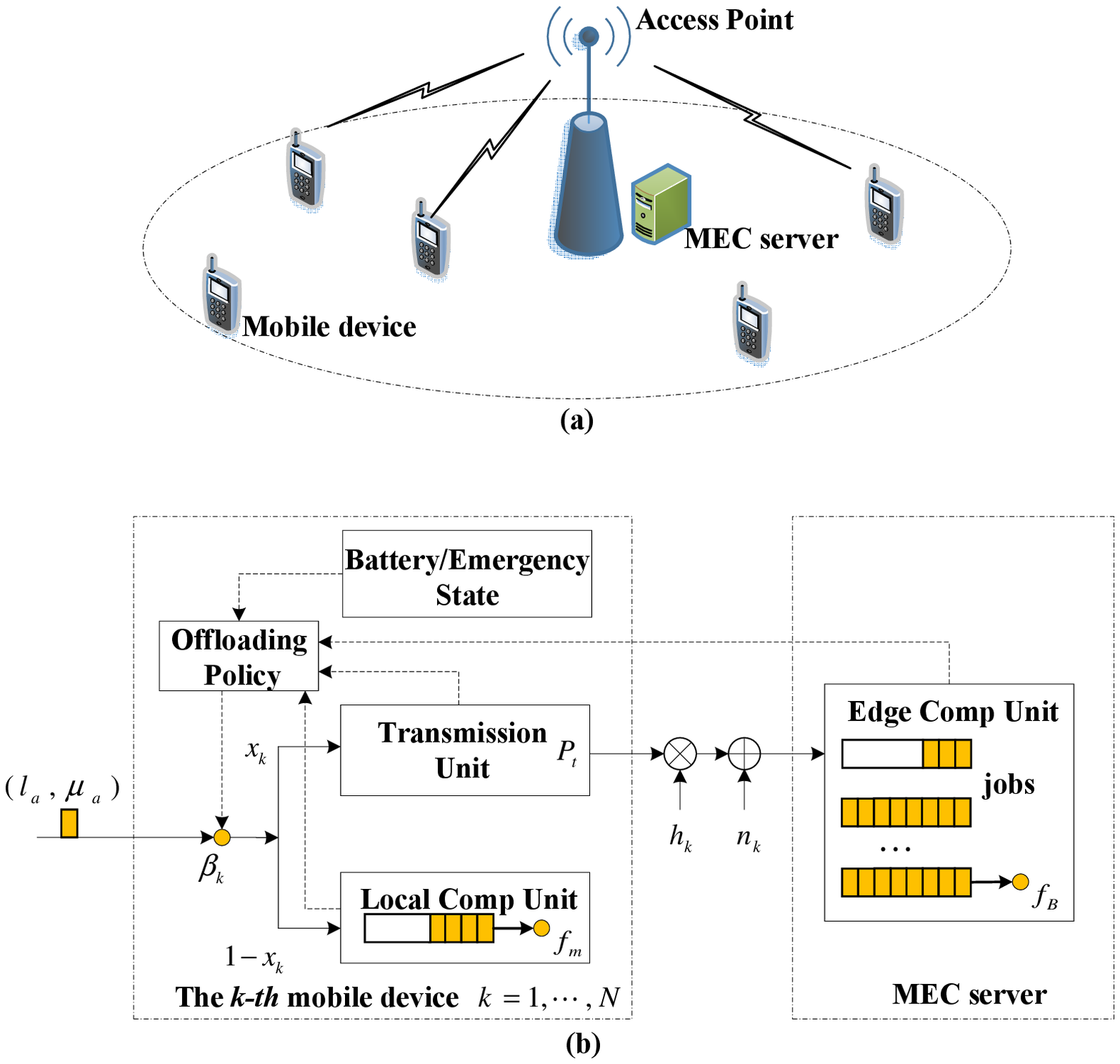}
 %where an .eps filename suffix will be assumed under latex,
% and a .pdf suffix will be assumed for pdflatex; or what has been declared via
\DeclareGraphicsExtensions. \caption{(a) A multi-user mobile-edge computing system.
(b) An illustration of the job arrival/offloading/computation model.}
\label{fig1}
\vspace* {-12pt}
\end{figure}

\section{System Model}
We consider an MEC system (see Fig. 1(a)) consisting of an AP and multiple mobile devices.
The wireless AP could be a small-cell base station, or a Wi-Fi AP.
Aside from serving as a conventional AP to the core network, it is installed with an additional
computing server and serves as an edge computing server. %, for the purpose of incentivising more users to join the system.
The mobile devices are running computation-intensive and delay-sensitive jobs, however,
they may have insufficient computing power or limited battery energy to complete those jobs.
As such, the devices that are short of computing or energy resources, should offload part of their
jobs to the AP. % for the purpose of completing jobs in time.
In what follows, we will introduce the offloading policy, the wireless channel model,
followed by the models for computing.%, followed by the economics model for offloading.

\subsection{The offloading policy and the radio access model}
Before proceeding to the offloading policy, we first introduce {\textit{the job generation model}}.
We assume that jobs arrive at the mobile device user following a Poisson process and at a rate of $\lambda_a$.
On the other hand, in wireless communication systems time is divided into slots having a unit duration of $t_0=1$ ms.
As such, we have a discrete-version of the Poisson arrival process, i.e., a Bernoulli process with the parameter $p_a=\lambda_at_0$ \cite{Refael03}.
The inter-arrival times at the mobile device are thus independent and i.i.d. random variables, following a
geometric distribution with the parameter $p_a$.
Besides, the service times are assumed to be identically distributed (i.i.d.) and exponentially distributed with parameter $\mu_a$.
In this paper, the job offloading is described by a tuple ($l_a, \mu_a$), where $\mu_a$ represents the average CPU cycles required to accomplish a job,
and the product $l_a\mu_a$ (e.g., the input parameters and the program codes) denotes the required offloading data size per job.

The mobile device users that are short of computing or energy resources, intend to offload all/part of their
jobs to the AP.
Specifically, when a job arrives, the mobile device will check its wireless channel condition to the AP.
We consider flat-fading channels and assume that the channel coherence time is
bigger than the time required to finish offloading a job.
As such, we assume that once a decision to offload its current job is made, the mobile device user %will get sufficient frequency bands such that it
could finish offloading in a channel block.
Hence, we consider the following \textit{offloading policy}.
When a job arrives and the channel is too
bad to carry data transmission, the mobile device user shall choose local computing, instead of
waiting for a favorable channel condition.
On the other hand, if the channel power gain is higher than
a threshold to support its expected transmission rate, denoted by $\beta_k$, the mobile device user would choose to
offload its job to the AP.

\begin{table}[t!]
  \begin{center}
    \caption{Summary of key notations}
    \label{tab:table1}
    \begin{tabular}{l|S}
      \toprule
      \textbf{Notation} & \textbf{Description} \\
      \hline
      $t_0$ & \textrm{The time duration per time slot} \\
      \hline
      $p_a$ & \textrm{Job arrival probability at the mobile device per time slot} \\
      \hline
      $\lambda_a$ & \textrm{Job arrival rate at the mobile device (jobs per second)} \\
      \hline
      $\mu_a$ & \textrm{Average CPU cycles needed by the computation job} \\
      \hline
      $l_a$ & \textrm{The input data size per cycle (in nat)} \\
      \hline
      $\kappa_m$ & \textrm{The energy coefficient of the mobile device} \\
      \hline
      $x_k$ & \textrm{User $k$'s offloading frequency} \\
      \hline
      $d_k$ & \textrm{User $k$'s distance to the access point} \\
      \hline
      $P_t$ & \textrm{User $k$'s transmit power} \\
      \hline
      $\sigma^2$ & \textrm{Received noise power at the access point} \\
      \hline
      $c_k^e$ & \textrm{User $k$'s weight of the computational energy} \\
      \hline
      $c_k^t$ & \textrm{User $k$'s weight of the computational time} \\
      \hline
      $h_k$ & \textrm{User $k$'s instantaneous small-scale channel gain} \\
      \hline
      $\beta_k$ & \textrm{User $k$'s transmission rate} \\
      \hline
      $f_m$ & \textrm{CPU-cycle frequency of the mobile device} \\
      \hline
      $\mu_m$ & \textrm{Computing service rate of User $k$ (jobs per second)} \\
      \hline
      $f_B$ & \textrm{CPU-cycle frequency of the MEC server} \\
       \hline
      $\mu_B$ & \textrm{Computing service rate of the MEC server (jobs per second)} \\
       \hline
      $g_k$ & \textrm{User $k$'s demand function} \\
      \hline
      $N$ & \textrm{The number of mobile device users} \\
      \bottomrule
    \end{tabular}
  \end{center}
\end{table}

%With the industry moving into the five-generation (5G) wireless communication era,
%an unprecedented spectrum and multi-Gigabit-per-second (Gbps) data rates are available to its users.
In this paper, we consider the scenario with sufficient wireless frequency bands, and
multiple mobile devices \textit{access to the AP in a FDMA mode}.
This is possible in the future five-generation (5G) wireless communication era, wherein
an unprecedented spectrum and multi-Gigabit-per-second (Gbps) data rates are available to its users.
In FDMA mode, each mobile device user is allocated a different frequency band, indicating that
they suffer no multi-user interference from each other.
Let $h_k$ denote the small-scale channel gain from the $k$-th mobile device to the AP, $k=1,2,\cdots, N$.
The achievable uplink data rate can thus be computed by
\begin{align}
R_k = {\rm log}(1+d_k^{-\alpha}|h_k|^2P_t/\sigma^2),\ k=1,2,\cdots, N,\label{eq1}
\end{align}
where $d_k$ denotes user $k$-th distance to the AP and $\alpha$ represents the path loss exponent.
$P_t$ is the transmission power and $\sigma^2$ denotes the received noise power at the AP.
Comparing the achievable data rate $R_k$ with the expected data rate $\beta_k$ and by Shannon theorem \cite{David05}, one can see that
when $R_k>\beta_k$ the mobile device could offload
its job to the AP successfully. By some mathematical derivations and letting $\rho_k=d_k^{-\alpha}P_t/\sigma^2$
be the received SNR, we arrive at the following conditions of successful offloading,
\begin{align}
|h_k|^2 > (e^{\beta_k}-1)\rho_k^{-1}, \  k=1,2,\cdots, N.\label{eq2}
\end{align}
These, combined with the aforementioned offloading policy, indicate that
the offloading frequencies (probabilities) are
\begin{align}
x_k={\rm Pr}(|h_k|^2 > (e^{\beta_k}-1)\rho_k^{-1}), \  k=1,2,\cdots, N. \label{eq3}
\end{align}

According to the expressions in (\ref{eq3}), it can be seen that
the offloading frequency at the mobile device user is a decreasing function of the threshold value of $\beta_k$.
As such, by adjusting $\beta_k$ the mobile device user is able to control the frequency it offloads jobs.

\subsection{Computation model}
According to the offloading policy,
%when a job arrives the mobile device user will check its channel condition.
%If its channel condition is good enough to support an expected data transmission rate of $\beta_k$,
%the mobile device user will offload its job to the AP;
jobs that are offloaded to the AP will be pushed into a buffer of the AP, awaiting for edge computing.
Otherwise, the jobs are pushed into
a local buffer, awaiting for local computing.
In what follows, we discuss the total overhead/cost in terms of local computing and edge computing.

\subsubsection{Local computing}
In the local computing approach the mobile device user executes the job
with its own computing power. Let $f_m$ be the mobile device's computing capability (CPU cycles
per second), which indicates that the service rate of local computing as $\mu_m t_0=(f_m/\mu_a)t_0$ (jobs per slot).
Thus, for each job the expected time spent in the system
(including both the computation execution time and the time awaiting in a local buffer)
can be given as
\begin{align}
D_k^{\rm LC}(x_k)&=\frac{1}{\mu_mt_0-p_a(1-x_k)}t_0 \nonumber \\
&\mathop {\rm{ = }}\limits^{(a)} \frac{1}{\mu_m-\lambda_a(1-x_k)}, \quad  k=1,2,\cdots, N,\label{eq4}
\end{align}
where (a) is due to $p_a=\lambda_at_0$.
The computational energy spent in local computing can be expressed as
\begin{align}
E_k^{\rm LC}&=\kappa_m f_m^2 \mu_a, \  k=1,2,\cdots, N, \label{eq5}
\end{align}
where $\kappa_m f_m^2$ is the power consumption per CPU cycle,
and $\kappa_m$ is an energy consumption coefficient that depends on the chip architecture \cite{Antti10}.

Combining (\ref{eq4}) and (\ref{eq5}) we can compute the total weighted cost by local computing as follows,
\begin{align}
Z_k^{\rm LC}(x_k)&=c_k^e E_k^{\rm LC}+c_k^t D_k^{\rm LC}(x_k), \  k=1,2,\cdots, N,\label{eq6}
\end{align}
where $0<c_k^e<1$ (in units 1/Joule) and $0<c_k^t<1$ (in units 1/Second) are the weights of the
computational energy and time.
With different weights $c_k^e$ and $c_k^t$ we allow different users to
put different emphasis in decision making.
For example, when the mobile device is at a low battery state,
it would tend to reduce the energy consumption in decision making.
As such, the user will choose a higher value of $c_k^e$.
On the other hand, when the user has a more stringent quality of experience
(QoE) requirement in delay, it would like to put more consideration to
the delay cost and it will set a bigger value of $c_k^t$.

\subsubsection{Edge computing}
In the edge computing approach the mobile device user will offload
its job to the AP and resort to the computing power of the AP.
First, it takes the mobile device some time to complete offloading, which can be
calculated as follows,
\begin{align}
D_{k,1}^{\rm EC}(x_k)&=\frac{l_a\mu_a}{\beta_k (x_k)}, \  k=1,2,\cdots, N.\label{eq7}
\end{align}
This, combined with the the fact each mobile device transmits with power $P_t$, indicates that the
energy required by offloading is
\begin{align}
E_k^{\rm EC}(x_k)&=P_t\frac{l_a\mu_a}{\beta_k (x_k)}, \  k=1,2,\cdots, N.\label{eq8}
\end{align}

On the other hand, it takes some time for the offloaded job to stay at the AP before it leaves after execution.
Based on the aforementioned offloading policy, the mobile device will choose to offload its jobs with probability $x_k$.
This, combined with the \textit{splitting} property of the queueing theory \cite{Refael03},
indicates that the job offloading from a mobile device user also follows a Poisson process, with parameter $\lambda_a x_k$.
Therefore, the job arrival at the AP is a superposition of multiple Poisson processes from multiple mobile devices, which, according to
the \textit{superposition} property of queueing theory \cite{Refael03}, is another Poisson process with the arrival rate as the
sum arrival rate of the superposed processes, i.e.,
$\sum\nolimits_{k=1}^{ N} \lambda_a x_k$.
In addition, the service times are i.i.d. and exponentially distributed with parameter $\mu_a$.
Therefore, we get a M/M/1 queue for computing at the AP.
Let $f_B$ be the AP's computing capability (CPU cycles
per second). The service rate of the mobile device is then $\mu_B=f_B/\mu_a$ (jobs per second).
Hence, for each job the expected time spent at the AP
(including both the computation execution time and the time spent awaiting in the edge buffer)
can be expressed as,
\begin{align}
D_{k,2}^{\rm EC}({\bf x})&=\frac{1}{\mu_B-\sum\nolimits_{k=1}^{ N} \lambda_a x_k}, \  \forall k,\label{eq9}
\end{align}
where ${\bf x} \triangleq(x_1, x_2, \cdots, x_N)$.

Similar to many existing studies such as \cite{Xuchen16,Fei18,Jianchao18}, we neglect the energy overhead of edge computing, due to
the fact that normally the AP can access to wired charing and it has no lack-of-energy issues.
As such, by combining (\ref{eq7}) to (\ref{eq9}) we could compute the total weighted cost by edge computing as follows,
\begin{align}
Z_k^{\rm EC}({\bf x})= c_k^e E_k^{\rm EC}(x_k)+c_k^t (D_{k,1}^{\rm EC}(x_k)+D_{k,2}^{\rm EC}({\bf x})),& \nonumber \\
   k=1,\cdots, N. &\label{eq10}
\end{align}
Note that the total cost by offloading as well as the end-to-end delay of computation
$D_{k,2}^{\rm EC}({\bf x})$ depend on both the local variable (i.e., $x_k$ for the $k$-th user) and also the offloading frequency of other users.
As we will see later, due to this coupled nature of delay we have coupled objective functions.

%In summary, in this paper we make the following assumptions.
%\begin{enumerate}
%\item The wireless channel is flat-fading and the channel correlation time is
%bigger than the time required to finish offloading a job. As such, we assume that
%once it chooses to offload, the mobile device user could finish offloading a job in a channel block.
%\item The channels associated with different users are independent of each other, and so is the
%offloading frequency of users. Hence, the arrival of
%of jobs at the AP also follows a Poisson process.
%\item The computing server executes the offloading jobs from users
%on a first-come, first-served (FCFS) basis.
%\end{enumerate}

\section{Game Formulation}
Due to the sharing nature of MEC server, the delay and hence the cost of offloading for users are coupled.
As such, it is usually difficult to jointly optimize users' offloading decisions centrally.
%In this section, we formulate problems based on game theoretic approaches, which aim to achieve
%efficient computation offloading decision makings among the mobile device users.
On the other hand, the mobile devices are owned by different individuals who may
have different QoE requirements on delay and pursue different interests.
In that case, the economics theory is a useful tool in dealing with interactions among users,
devising decentralized mechanisms with low complexity, and inducing users
to self-organize into a mutually satisfactory solution.

In the following, we will formulate games from the perspective of economics.
Towards that goal, we first introduce a business model for mobile edge computing, which includes physical layer parameters in
the utility function and the cost function. Then, based on the proposed business model we formulate two games,
with one maximizing the individual's interest and the other maximizing the overall system's interest.

\subsection{Proposed economics model for mobile edge computing}
By the aforementioned offloading policy, upon the arrival of a new job, the mobile device will
choose to offload with probability $x_k$, % which is a function of a predefined threshold $\beta_k$;
or push it into a local buffer for local computing with probability $\bar{x}_k$ ($\bar{x}_k \triangleq1-x_k$).
Therefore, the expected total cost can be written as follows,
\begin{align}
Z_k({\bf x}) &= \bar{x}_k Z_k^{\rm LC}({x_k}) + x_k  Z_k^{\rm EC}({\bf x}). \label{eq11}
%   &= (c_k^e E_k^{\rm LC}+c_k^t D_k^{\rm LC})(1-x_k)+ (c_k^e E_k^{\rm EC}+c_k^t (D_{k,1}^{\rm EC}+D_{k,2}^{\rm EC}))x_k \label{eq11a}
\end{align}

On the other hand, when there is no such edge server providing computing power, users have
to complete every job by themselves and the average cost running a job locally is
$Z_k^{\rm LC}(0)$.
This, combined with (\ref{eq11}), indicates that the gross profit of
offloading by the $k$-th user under a given offloading strategy ${\bf x}$ is,
\begin{align}
V_k({\bf x}) &=Z_k^{\rm LC}(0)-Z_k({\bf x}), \  k=1, \cdots, N. \label{eq12a}
\end{align}

The key idea of the business model is to introduce the utility function and the cost function.
Several observations are in order. Firstly, the profit each user obtains equals the cost savings
from offloading, and it is a linear combination of the
energy costs and the delay costs.
Secondly, the coupled delay cost $D_{k,2}^{\rm EC}({\bf x})$ reflects the harm/congestion each user
causes to the other users, with a bigger value indicating that the AP provides worse service to the users.
Thirdly, except for the expected time spent at the AP, i.e., $D_{k,2}^{\rm EC}({\bf x})$, which depends
on all users' offloading decisions, the other items in the profit function
only depend on each user's own offloading frequency $x_k$.
Motivated by these observations, we introduce a utility function $U_k(x_k)$ which includes the
items in the profit function that only depend on the local variable $x_k$, i.e.,
\begin{align}
U_k(x_k)=&Z_k^{\rm LC}(0)-\bar{x}_k Z_k^{\rm LC}({x_k}) \nonumber \\
&- x_k(c_k^e E_k^{\rm EC}(x_k)+c_k^t D_{k,1}^{\rm EC}(x_k)).  \label{eq13}
\end{align}
This, combined with (\ref{eq12a}), indicates that
\begin{align}
V_k({\bf x})=U_k(x_k)-C_k({\bf x}), \  k=1, \cdots, N,   \label{eq12b}
\end{align}
where $C_k({\bf x}) \triangleq c_k^t x_k D_{k,2}^{\rm EC}({\bf x})$ can be regarded as the cost due to the
sharing of an MEC server at the AP.

In economics, the utility function measures the welfare a consumer obtains, as a function of the consumption of
real goods such as food, clothes, and so on. Here, we apply a utility function to capture
the benefit a user can obtain by offloading. The offloading frequency $x_k$ quantifies the amount a user is willing to
offload; a higher offloading frequency indicates that the user is more willing to offload.
By comparing the utility function and the cost function in (\ref{eq12b}), each user could
find its indifferent point of offloading frequency, i.e., the point at which the user achieves a maximum profit.
It turns out that the utility function in (\ref{eq13}) is strictly concave and strictly increases with respect to users' offloading frequency,
satisfying the law of diminishing marginal returns in economics;
%thus bridging the gap between the economics and physical layer parameter optimizations;
this is discussed in the following lemmas.

\begin{lemma}\label{lemma1}
\textit{Taking the derivative of $U_k(x_k)$ with respect to $x_k$, we arrive at
the demand function of user $k$, i.e.,
\begin{align}
&g_k({x}_{k})\triangleq  \dfrac{\partial U_k(x_k)}{\partial x_k}. \nonumber
\end{align}
For each mobile device user, it holds true that the
demand function $g_k({x}_{k})$ is a monotonically decreasing function of the offloading frequency $x_k$.
Moreover, there exists a unique solution of the equation $g_k({x}_{k})=0$, denoted as $x_k^{up}$. }
\end{lemma}
\begin{IEEEproof}
See Appendix A%\ref{appA}
\end{IEEEproof}
\medskip
\emph{Lemma} \ref{lemma1} indicates that
the utility function is a concave function of the offloading frequency $x_k$.

\medskip
\begin{lemma}\label{lemma2}
\textit{For each mobile device user, it holds true that $U_k(0)=0$, and the utility function $U_k(x_k)$ is a
monotonically increasing function of the offloading frequency $x_k \in [0, x_k^{up}] $.   }
\end{lemma}
\begin{IEEEproof}
By the definition of $U_k(x_k)$, it is easy to verify that $U_k(0)=0$.
On the other hand, according to \textit{Lemma 1}, it is clear that $g_k(x_k^{up})=0$.
In addition, since $g_k({x}_{k})$ is a monotonically decreasing function of $x_k$,
it holds true that $g_k({x}_{k})>0$ when $x_k< x_k^{up}$.
This completes the proof.
\end{IEEEproof}
\medskip

\begin{lemma}\label{lemma3}
\textit{If the $i$-th mobile device user is closer to the AP than the $j$-th user, that is, $d_i<d_j$,
then its utility function strictly dominates that of the $j$-th user, %it has a bigger utility function,
i.e., $U_i(x) > U_j(x)$.   }
\end{lemma}
\begin{IEEEproof}
The proof is omitted since it could be easily verified with the formula of $U_k(x_k)$ in (\ref{eq13}).
\end{IEEEproof}

\begin{figure}[!h]
\centering
\includegraphics[width=3in]{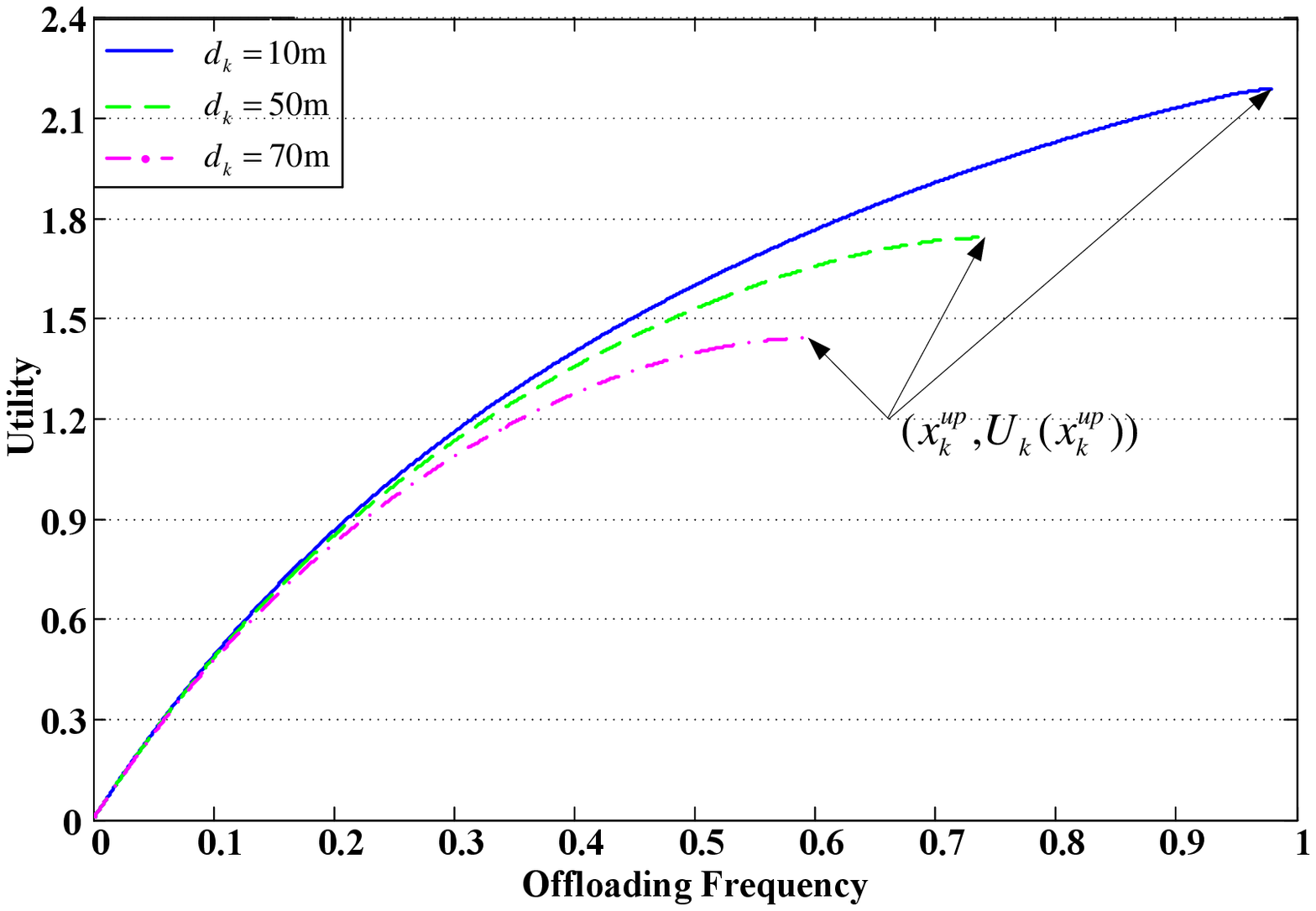}
 %where an .eps filename suffix will be assumed under latex,
% and a .pdf suffix will be assumed for pdflatex; or what has been declared via
\DeclareGraphicsExtensions. \caption{The achievavble utility for varying Mobile-AP distances.}
\label{fig0}
%\vspace* {-6pt}
\end{figure}

\emph{Lemma} \ref{lemma1} to \ref{lemma3} provide the properties of the utility curve.
That curve is illustrated in Fig. \ref{fig0} (as a function of the offloading frequency), for
a system with $c_k^t=0.9$, $c_k^e=0.1$, $f_m=0.1$GHz. The job arrives at a rate of $\lambda_a=0.6$ jobs per second, with each job
requiring an average of $\mu_a=100$M CPU-cycles to run and an average of $l_a\mu_a=100$ nats to offload.
Different distances, i.e., $d=10$m, $d=50$m and $d=70$m, from the user to the AP are respectively considered.
One can see that the utility function is strictly increasing but the increasing rate (i.e., the demand function of $g(x_k)$)
decreases as the offloading frequency increases; this is consistent with the law of diminishing marginal returns in economics, i.e.,
the more the user consumes, the demand for additional unit of goods decreases.
Moreover, the user closer to the AP has a greater demand to offload and can achieve a higher utility value with the same offloading amount.
This is keeping with the intuition that the closer the user is to the AP, the better the wireless channel it experiences.

\subsection{Game formulations}
We first consider a decentralized computation offloading decision making problem,
wherein for any given strategies of others, the mobile device user intends to
maximize its own profit $V_k({\bf x})$ by optimizing the local offloading frequency $x_k$ via adjusting
its threshold parameter $\beta_k$.
Specifically, we have the following optimization problem that each mobile device user needs to solve for.

\underline{\textbf{Problem 1 (Selfish Problem):}}
\begin{align}
x_k^{\star} \triangleq &{\arg}  \max_{x_k} \quad  V_k({\bf x}) \nonumber \\
& {\rm s.t.} \quad  0 \le x_k  \le 1. \label{eq14}
\end{align}

\textit{Definition 1: A strategy profile ${\bf x}^\star =(x_1^\star, \cdots, x_N^\star)$ is
a Nash Equilibrium (NE) of the offloading game if at the equilibrium ${\bf x}^\star$, no player can further increase
its profit by unilaterally changing its strategy, which indicates that
\begin{align}
V_k({x}_{k}^\star,{\bf x}_{-k}^\star) \ge  V_k({x}_{k},{\bf x}_{-k}^\star),  \ \forall  {x}_{k} , k=1, \cdots, N. \label{eq15}
\end{align}
where ${\bf x}_{-k}^\star \triangleq (x_1^\star, \cdots, {x}_{k-1}^\star, {x}_{k+1}^\star, \cdots, x_N^\star)$.
}
\medskip

With Problem 1, we formulate a game in which each mobile device user is selfish and responses
to maximize its own benefit. The objective of this game is to find the NE point from which
no mobile device user has incentives to deviate.
However, the solution of Problem 1 is locally optimal; it may not coverage to a point where the sum
profit of all the users is maximized.

On the other hand, we consider a social-welfare computation offloading decision making problem.
Assume that the AP acts as a social planner, usually referred to as the hand of God in economics.
It would like users to choose their offloading decisions
such that the sum profit $\sum\nolimits_{k=1}^{ N} V_k({\bf x})$ is maximized.
%Exact knowledge of users' utility functions are assumed to be available at the AP,
%which could be obtained by asking users to report about their local information.
%to induce users to choose the right response balancing between its own profit and the harm it causes to the other users.
%Specifically, for any given strategies of others, the mobile device user optimizes its offloading decision $x_k$  to
%maximize the sum profit $\sum\nolimits_{k=1}^{ N} V_k({\bf x})$, via adjusting
%its threshold parameter $\beta_k$.
Specifically, we have the following optimization problem that each mobile device user needs to solve for.

\underline{\textbf{Problem 2 (Social Problem):}}
\begin{align}
{ \bar x}_k^{\star} \triangleq &{\arg}  \max_{ x_k} \quad \sum\nolimits_{k=1}^{ N} V_k({\bf x}) \nonumber \\
& {\rm s.t.} \quad  0 \le x_k  \le 1. \label{eq16}
\end{align}

\textit{Definition 2: A strategy profile ${\bf \bar x}^\star =(\bar x_1^\star, \cdots, \bar x_N^\star)$ is
a Social Equilibrium (SE) of the offloading game if at the equilibrium ${\bf \bar  x}^\star$, no player can further increase
the sum profit of the whole system (social welfare) by unilaterally changing its strategy, which indicates that
\begin{align}
\sum\nolimits_{k=1}^{ N} V_k({\bar x}_{k}^\star,{\bf \bar x}_{-k}^\star) \ge  \sum\nolimits_{k=1}^{ N} V_k({  x}_{k},{\bf \bar x}_{-k}^\star), \ \forall {x}_{k} . \label{eq17}
\end{align}
where $\bar{\bf x}_{-k}^\star \triangleq (\bar x_1^\star, \cdots, \bar{x}_{k-1}^\star, \bar{x}_{k+1}^\star, \cdots, \bar x_N^\star)$.
}
%\medskip

Intuitively, in Problem 2 each user should also be concerned with the congestion it causes to other users
and should keep its offloading under an appropriate amount for other users' welfare;
the difficulty lies in how to incentivise users to do so when
users are selfish and will choose their offloading decisions such that its
individual profit $V_k(x_k)$ is maximized.

Pricing is a useful tool in incentivising users to choose the socially optimal levels of demand.
The key idea is to enforce users to pay for the congestion it causes to the other users.
In the following Problem 3, we study the pricing-based scheme, which
charges users an additional edge computing service fee to regulate users' behavior.

\underline{\textbf{Problem 3 (Regulated Selfish Problem):}}
\begin{align}
{\hat x}_k^{\star} \triangleq &{\arg}  \max_{x_k} \quad  V_k({\bf x}) -Px_k \nonumber \\
& {\rm s.t.} \quad  0 \le x_k  \le 1, \label{eq01}
\end{align}
where $P$ denotes the unit price for offloading.
In the following section, our goal is to derive an optimal price of $P$ such that
the interactive decision-making process with regulated self-interested users
converges to a Nash Equilibrium point equal to the Social Equilibrium point.
It is worth noting that in a practical system the optimal pricing could be learned based on the historical supply-demand relationship.
Due to limitations of space, in this paper we only give discussions on the existence of
such optimal price and show the advantages it brings in terms of the overall system performance.

%As to the detailed implementation of the VCG mechanism, please refer to \cite{Courcoubetis03}.
%the price is applied to force users to take into account the negative effects of congestion they cause to other users.

\section{Optimal Solutions of Offloading}
%By definition, one can see that in the game of Problem 1 each user
%optimizes its offloading frequency to maximize its individual interest, while in the
%game of Problem 2 each user adjusts its offloading frequency to maximize the overall system performance.
%Intuitively, in the latter game each user is also concerned with the congestion it causes to the other users
%and will keep its offloading under an appropriate amount for the other users' welfare.
%As such, for the purpose of improving the overall system performance, it is critical to regulate users'
%behaviors appropriately and efficiently. This could be done by the pricing-based idea from the economics, i.e.,
%imposing additional price for providing the edge computing service, through which the regulator
%ensures the market operating efficiently since it allocates goods/services amongst consumers so that they go to people who value them the most.

In this section, we aim to derive an optimal price to regulate users' behavior, such that by this pricing-based scheme
the individual objectives of self-interested users are aligned with that of the overall system.
Towards that goal, we first analyze the structural property of the games and admit in closed form
the Nash Equilibrium and the Social Equilibrium, respectively. Based on these results, we then
propose the optimal price. Under this pricing-based scheme,
the interactive decision-making process with self-interested users
converges to a Nash Equilibrium point equal to the Social Equilibrium point.
In this way, we provide an algorithm which improves the overall resource efficiency
and enhances the system performance in a distributed way.

For the purpose of getting more insights into the offloading problem, we first investigate
the case in which users have the same time cost weight as well
as the same energy cost weight, i.e., $c_k^t=c_0^t$, $c_k^e=c_0^e$, $\forall k$.
%In this case and by definition, the mobile device users have the same demand function and the same utility function.
%For the ease of exposition, in this section, we denote by $g_k=g_0$, $\forall k$.

\subsection{Analysis of Nash Equilibrium}
The key idea for deriving the equilibrium is to look into the necessary conditions,
i.e., at the equilibrium the derivative of the objective function should be zero.

We first look into Problem 1. At the equilibrium it holds that
\begin{align}
&\dfrac{\partial U_k(x_k)}{\partial x_k}- \dfrac{\partial  x_k c_k^t D_{k,2}^{\rm EC}({\bf x})}{\partial x_k}=0 \nonumber \\
\Leftrightarrow& g_k(x_k)= \frac{c_0^t (\mu_B-\sum\nolimits_{j \ne k}^{ N} \lambda_a x_j)}{(\mu_B-\sum\nolimits_{j=1}^{ N} \lambda_a x_j)^2},\ k=1, \cdots, N. \label{eq18}
\end{align}

Combing those $N$ equations in (\ref{eq18}) yields the following Theorem.
\medskip
\begin{theorem}\label{theorem1}
\textit{For the case with users of the same demand function, denoted by $g_k=g_0$, $\forall k$ (also referred to
as the homogeneous user case in the following), %i.e., $c_k^t=c_0^t$, $c_k^e=c_0^e$, $\forall k$,
at the Nash Equilibrium each mobile device user has
the same value of offloading frequency $x_0^\star$, which can be achieved via setting
a same parameter of $\beta_k^\star$. Moreover, the offloading frequency satisfies the following equation,
\begin{align}
N\lambda_a x_0^\star +\sqrt{\dfrac{c_0^t(\mu_B-(N-1)\lambda_a x_0^\star)}{g_0(x_0^\star)}}=\mu_B.  \label{eq19}
\end{align}
}
\end{theorem}
\begin{IEEEproof}
See Appendix B%\ref{appB}
\end{IEEEproof}
\medskip

\begin{corollary}\label{corollary1}
\textit{For homogeneous user case, if the following equation is satisfied,
\begin{align}
c_0^e  E_k^{\rm LC} + \dfrac{c_0^t \mu_m}{(\mu_m-\lambda_a)^2}>\dfrac {c_0^t}{\mu_B},  \label{eq20}
\end{align}
there exists a positive unique Nash Equilibrium.
Otherwise, no mobile device would like to offload, which corresponds to a
trivial Nash Equilibrium, i.e., $x_k^\star=0$, $\forall k$. }
\end{corollary}
\begin{IEEEproof}
See Appendix C%\ref{appC}
\end{IEEEproof}
\medskip

\subsection{Analysis of Social Equilibrium}
In Problem 2 the users should also take into account the negative effects of congestions it causes to the other users.
As such, at the equilibrium it holds that
\begin{align}
&\dfrac{\partial U_k(x_k)}{\partial x_k}- \sum\nolimits_{j=1}^{ N}\dfrac{\partial c_0^t x_j D_{j,2}^{\rm EC}({\bf x})}{\partial x_k}=0 \nonumber \\
\Leftrightarrow& g_k(x_k)= \frac{c_0^t \mu_B}{(\mu_B-\sum\nolimits_{j=1}^{ N} \lambda_a x_j)^2},\ k=1, \cdots, N, \nonumber \\
\Leftrightarrow& \mu_B-\sum\nolimits_{j=1}^{ N} \lambda_a x_j = \sqrt{\dfrac{{c_0^t}\mu_B}{ g_k (x_k)}},\ k=1, \cdots, N. \label{eq21}
\end{align}

%By some algebra derivations of (\ref{eq21}), we arrive at
%\begin{align}
%\mu_B-\sum\limits_{j\ne k}^{ N} \lambda_a x_j = F_S(x_k),\ k=1, \cdots, N,  \label{eq21b}
%\end{align}
%where $F_S(x_k)\triangleq\lambda_a x_k+\sqrt{{{c_0^t}\mu_B}/{ g_k (x_k)}}$.
%
%The equations in (\ref{eq21b}) give the necessary conditions of the equilibrium of the game, indicating
%the best response of a user when the other users' offloading decisions are given.
%Based on (\ref{eq21b}) we update the offloading frequency $x_k$ iteratively.
%We summarize this decentralized pricing-based algorithm as in Table Algorithm \ref{alg:SE}.
%
%\begin{algorithm}
%\caption{Social-optimal distributed algorithm}\label{alg:SE}
%\begin{algorithmic}[1]
%%\Procedure{Euclid}{$a,b$}\Comment{The g.c.d. of a and b}
%\State \textbf{Initialization}: $x_k^0=0$, $\forall k$
%\State $t\gets 0$
%\While{$ \delta_x > \epsilon$}   \Comment{We have the answer if $ \delta_x \le \epsilon$}
%%\State $x_k^{temp} \gets x_k^{t+1}, \forall k$
%\For{$k=1, \cdots, N$} %\Comment{We have the answer if r is 0}
%\State $b \gets \mu_B-\sum\nolimits_{j=1}^{k-1} \lambda_a x_j^{t+1} -\sum\nolimits_{j=k+1}^{N} \lambda_a x_j^{t}$
%\State $x_k^{t+1} \gets$ Solve $F_S(x_k^{t+1})=b$ for $x_k^{t+1}$
%\EndFor\label{euclidendfor}
%\State  \textbf{end}
%\State $\delta_x \gets  \frac{1}{ N} \sum\nolimits_{k=1}^{N}|x_k^{t+1}-x_k^{t}|$
%\State $t \gets t+1$
%\EndWhile\label{euclidendwhile}
%\State \textbf{end}
%\State \textbf{return} $\beta_k(x_k)$, $\forall k$  \Comment{According to (\ref{eq3})}
%%\EndProcedure
%\end{algorithmic}
%\end{algorithm}

The equations in (\ref{eq21}) give the necessary conditions of the social equilibrium by Problem 2.
Based on (\ref{eq21}) we discuss the existence as well as the uniqueness of the Social Equilibrium,
which is summarized in the following theorem.

\medskip
\begin{theorem}\label{theoremA}
\textit{There exists a unique Social Equilibrium point, at which the sum profit of users
is maximized and no user can further increase
the sum profit by unilaterally changing its strategy.
}
\end{theorem}
\begin{IEEEproof}
See Appendix D%\ref{appD}
\end{IEEEproof}
\medskip

Furthermore, for the homogeneous user case, the Social Equilibrium point can be derived in closed-form,
which is summarized in the following theorem.
\medskip
\begin{theorem}\label{theorem2}
\textit{For the homogeneous user case, %i.e., $c_k^t=c_0^t$, $c_k^e=c_0^e$, $\forall k$,
at the Social Equilibrium each mobile device user has the same value of offloading frequency $\bar x_0^\star$,
which can be obtained via setting a same parameter of $\bar \beta_k^\star$.
Moreover, the offloading frequency satisfies the following equation,
\begin{align}
N\lambda_a \bar x_0^\star +\sqrt{\dfrac{c_0^t\mu_B}{g_0(\bar x_0^\star)}}=\mu_B.  \label{eq22}
\end{align}
}
\end{theorem}
\begin{IEEEproof}
For any given offloading frequency vector ${\bf x}=(x_1, \cdots, x_N)$ satisfying (\ref{eq21}),
it holds that $g_0(x_m)=g_0(x_n)$, $\forall m, n$.
Moreover, according to \textit{Lemma} \ref{lemma1}, $g_0(x)$ is a monotonically decreasing function of $x$.
Therefore, $x_m=x_n=\bar x_0^\star$, $\forall m, n$.
Substituting this into (\ref{eq21}) yields (\ref{eq22}).
This completes the proof.
\end{IEEEproof}
\medskip

\begin{corollary}\label{corollary2}
\textit{For the homogeneous user case, if the following inequality is satisfied,
\begin{align}
c_0^e  E_k^{\rm LC} + \dfrac{c_0^t \mu_m}{(\mu_m-\lambda_a)^2}>\dfrac {c_0^t}{\mu_B},  \nonumber
\end{align}
there exists a positive unique Social Equilibrium.
Otherwise, no mobile device would like to offload, which corresponds to a
trivial Social Equilibrium, i.e., $\bar x_k^\star=0$, $\forall k$. }
\end{corollary}
\begin{IEEEproof}
The proof is omitted since it is similar to that of \textit{Corollary} \ref{corollary1}.
\end{IEEEproof}
\medskip

\begin{corollary}\label{corollary3}
\textit{Comparing the Nash Equilibrium and the Social Equilibrium, the
former converges to a Equilibrium point with a higher offloading frequency, i.e.,
\begin{align}
x_0^\star > \bar x_0^\star.  \nonumber
\end{align}
}
\end{corollary}
\begin{IEEEproof}
Clearly, it holds that $c_0^t(\mu_B-(N-1)\lambda_a x_0^\star) < c_0^t\mu_B$, which,
combined with (\ref{eq19}) and (\ref{eq22}), indicates that
$g_0(x_0^\star) < g_0(\bar x_0^\star)$. Besides, according to \textit{Lemma} \ref{lemma1},
$g_0(x)$ is a monotonically deceasing function. As such, we have $x_0^\star > \bar x_0^\star$.
This completes the proof.
\end{IEEEproof}
\medskip
\emph{Corollary} \ref{corollary3} is consistent with the intuition that
less jobs will be offloaded by the users who aim to solve for Problem 2, since they
put additional consideration on the harm offloading causes to the other mobile device users.

\subsection{Optimal pricing-based scheme to achieve Social Equilibrium}
According to \emph{Corollary} \ref{corollary3}, less jobs will be offloaded by users
for the purpose of maximizing the sum profit of users.
However, in the market every user is selfish and aims to maximize its own benefit.

In this subsection we study the regulated selfish problem, which regulates users' action via charging
them appropriate economic expenses for providing the edge computing service.
%Specifically, let $P$ denote
%the unit price for offloading, then the gross profit of each mobile device user by offloading is
%\begin{align}
%\hat V_k(x_k)=V_k(x_k)-P x_k,\ k=1, \cdots, N.  \label{eq23}
%\end{align}
The goal is to align the individual objectives of self-interested users with that of the overall system,
such that the interactive decision-making process with self-interested users
converges to a Nash Equilibrium point equal to the Social Equilibrium point.

Recall Problem 3 where each mobile device user is selfish and aims to maximize its own benefit.
The user is required to pay for offloading at a unit price of $P$.
As such, we replace $V_k(x_k)$ in Problem 1 with $\hat V_k(x_k)=V_k(x_k)-Px_k$.
Taking the derivative of $\hat V_k(x_k)$ with respect to $x_k$ and letting it be zero, we arrive at
similar formulas as in (\ref{eq18}), i.e.,
\begin{align}
g_k(x_k)-P= \frac{c_0^t (\mu_B-\sum\nolimits_{j \ne k}^{ N} \lambda_a x_j)}{(\mu_B-\sum\nolimits_{j=1}^{ N} \lambda_a x_j)^2},\ k=1, \cdots, N. \label{eqP1}
\end{align}

Letting $\hat g_k (x_k)=g_k(x_k) -P$ and by some algebra derivations of the equations in (\ref{eqP1}), we obtain
\begin{align}
\mu_B-\sum\limits_{j\ne k}^{ N} \lambda_a x_j = F_N(x_k),\ k=1, \cdots, N,  \label{eq26}
\end{align}
where $F_N(x_k)\triangleq\lambda_a x_k+\dfrac{c_0^t}{2\hat g_k (x_k)}+\sqrt{(\dfrac{{c_0^t}}{2\hat g_k (x_k)})^2+\dfrac{c_0^t\lambda_a x_k}{\hat g_k (x_k)}}$.
%\begin{align}
%F(x_k)=\lambda_a x_k+\frac{c_0^t}{2\hat g_0 (x_k)}+&\sqrt{(\frac{{c_0^t}}{2\hat g_0 (x_k)})^2+\frac{c_0^t\lambda_a x_k}{\hat g_0 (x_k)}}.  \label{eq27}
%\end{align}

\begin{algorithm}
\caption{Pricing-based social-optimal algorithm}\label{alg:NE}
\begin{algorithmic}[1]
%\Procedure{Euclid}{$a,b$}\Comment{The g.c.d. of a and b}
\State \textbf{Initialization}: $x_k^0=0$, $\forall k$
\State $t\gets 0$
\While{$ \delta_x > \epsilon$}   \Comment{We have the answer if $ \delta_x \le \epsilon$}
%\State $x_k^{temp} \gets x_k^{t+1}, \forall k$
\For{$k=1, \cdots, N$} %\Comment{We have the answer if r is 0}
\State $b \gets \mu_B-\sum\nolimits_{j=1}^{k-1} \lambda_a x_j^{t+1} -\sum\nolimits_{j=k+1}^{N} \lambda_a x_j^{t}$
\State $x_k^{t+1} \gets$ Solve $F_N(x_k^{t+1})=b$ for $x_k^{t+1}$
\EndFor\label{euclidendfor}
\State  \textbf{end}
\State $\delta_x \gets  \frac{1}{ N} \sum\nolimits_{k=1}^{N}|x_k^{t+1}-x_k^{t}|$
\State $t \gets t+1$
\EndWhile\label{euclidendwhile}
\State \textbf{end}
\State \textbf{return} $\beta_k(x_k)$, $\forall k$  \Comment{According to (\ref{eq3})}
%\EndProcedure
\end{algorithmic}
\end{algorithm}

The equations in (\ref{eq26}) give the necessary conditions of the equilibrium of the game, indicating
the best response of a user when the other users' offloading decisions are given.
Based on (\ref{eq26}) we update the offloading frequency $x_k$ iteratively.
The iteration stops until the offloading frequency difference of two neighbor iterations almost equals zero,
which indicates that we arrive at the NE point.
We summarize this decentralized pricing-based algorithm as in Table Algorithm \ref{alg:NE}.
Combing \textit{Theorem} \ref{theoremA} and \textit{Theorem} \ref{theorem3} indicates that
for the homogeneous user case we arrive at a NE equal to the unique SE.
In addition, as we will show later in the numerical results part,
for the heterogeneous user case this decentralized pricing-based algorithm
also converges.

It is worth noting that in a practical system and for the purpose of determining the best response, instead of all the other users'
offloading decisions, a user only needs to get to know the left hand side of Eq. (\ref{eq26})
which could be learned from its historical delays of edge computing.
From this point, an online user-behavior learning problem for the evolutional offloading strategy
is attracting, which is one of the our extension works in the near future.

By far, we have finished deriving a framework for determining the offloading strategy.
What remains is to determine the optimal price such that the proposed
distributed algorithm converges to a Nash Equilibrium equal to the Social Equilibrium.
On comparing the equations in (\ref{eq21}) and (\ref{eqP1}), for the
purpose of obtaining a Nash Equilibrium equal to the Social Equilibrium,
we should set the price as follows,
\begin{align}
P= \frac{c_0^t (\sum\nolimits_{j \ne k}^{ N} \lambda_a \bar x_j^\star)}{(\mu_B-\sum\nolimits_{j=1}^{ N} \lambda_a \bar x_j^\star)^2}
\thickapprox \frac{c_0^t (\sum\nolimits_{j =1}^{ N} \lambda_a \bar x_j^\star)}{(\mu_B-\sum\nolimits_{j=1}^{ N} \lambda_a \bar x_j^\star)^2}, \label{eq27}
\end{align}
where the approximation is due to the intuition of applying a uniform pricing
since users are setting the same weights of computational energy of $c_0^t$.
It should be expected that the approximation in (\ref{eq27}) is good when there are a large number of users.

Meanwhile, we consider the special homogeneous user case, in which
the optimal price could be derived in closed-form. Specifically,
substituting (\ref{eqP1}) into \emph{Theorem} \ref{theorem1} and comparing the result with \emph{Theorem} \ref{theorem2},
we arrive at \emph{Theorem} \ref{theorem3} as follows.

\medskip
\begin{theorem}\label{theorem3}
\textit{Considering the homogeneous user case, for the Nash Equilibrium point to equal to the Social Equilibrium point, the price should be set as
\begin{align}
P= \dfrac{(N-1)\lambda_a}{\mu_B}{\bar x_0^\star g_0(\bar x_0^\star)}.  \nonumber
\end{align}
}
\end{theorem}
\begin{IEEEproof}
See Appendix E%\ref{appE}
\end{IEEEproof}

%For the heterogeneous user case where users have different demand functions,
%it is difficult to derive a rigious optimal price. Meanwhile, it can be seen later in the simulation results that,
%by setting the edge computing service fee as
%\begin{align}
%P= \dfrac{(N-1)\lambda_a }{\mu_B} \dfrac{1}{N} \sum\nolimits_{k=1}^{N}\bar x_k^\star g_k(\bar x_k^\star). \label{eq27}
%\end{align}
%the interactive decision-making decision process converges to a Nash Equilibrium equal to the Social Equilibrium.

\section{Numerical Results}
In this section, we conduct simulations to validate our theoretical findings and the proposed algorithm (see Algorithm 1)
for both the homogeneous and the heterogeneous user cases.
We consider a system model as illustrated in Fig. 1. %simple semi-symmetric
For simplicity, we consider the symmetric case where $c_k^t=c_0^t=0.9$, $c_k^e=c_0^e=0.1$, $\forall k$.
The computing power at the AP and the mobile device users are $f_B=3$GHz and $f_m=0.1$GHz, respectively \cite{Antti10}.
The job arrives at the mobile device user at a rate of $\lambda_a=0.6$, with each job
requiring an average of $\mu_a=100$M CPU-cycles to run and $l_a\mu_a=100$ nats of input data size to offload.

The channels are modeled as multipath flat fading. The effect of the
channel between any transmit-receive pair on the transmitted signal
is modeled by a multiplicative scalar of the form $d^{-\alpha} h_k$ \cite{Inaltekin09}, where
$d$ is the distance between the two terminals, $\alpha=3.5$ is the path loss exponent.
The small-scale channel gains are assumed to be identically distributed (i.i.d.) and exponentially distributed with parameter $1$,
i.e., $|h_k|^2 \sim {\rm exp}(1)$. Substituting this distribution into (\ref{eq3}) yields
\begin{align}
x_k={\rm exp}(-(e^{\beta_k}-1)\rho_k^{-1}), \  k=1,2,\cdots, N. \nonumber
\end{align}

The transmit power is $P=100$mW and the noise power level is $\sigma^2=-40$dBm.
Unless otherwise specified, we set $d=50$m. As such, the received SNR at the
AP is $\rho_k=d_k^{-\alpha}P_t/\sigma^2=0.89$.
The stop threshold $\epsilon = 10^{-3}$.

\subsection{The special homogeneous user case}
In this subsection, we consider the homogeneous scenario in which users are uniformly distributed on
a circle of radius $R=d$ (unit: meters) and the AP locates at the center;
in this case, all the users have the same utility functions.

\begin{figure}[!h]
\centering
\includegraphics[width=3in]{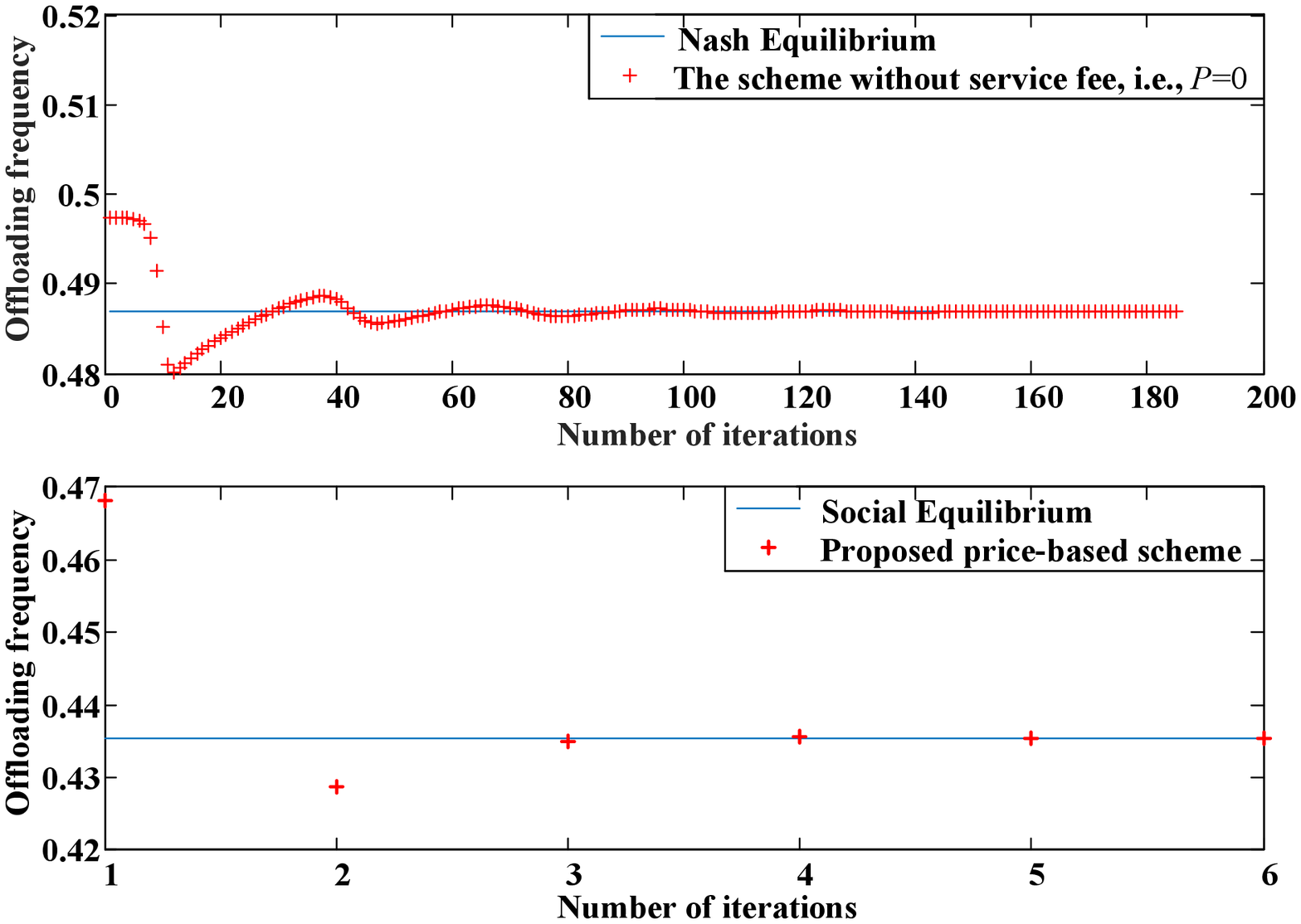}
 %where an .eps filename suffix will be assumed under latex,
% and a .pdf suffix will be assumed for pdflatex; or what has been declared via
\DeclareGraphicsExtensions. \caption{Convergence of the proposed algorithm in the homogeneous user case.}
\label{fig2}
\vspace* {-6pt}
\end{figure}

Fig. \ref{fig2} illustrates the convergence of the proposed algorithm.
The scheme with the price $P=0$ and the proposed scheme (see Table Algorithm 1)
with the optimal price (see \emph{Theorem} \ref{theorem3}) are considered respectively.
The local information of $F_N(x_k)$ and the edge delay-related information $\mu_B-\sum\nolimits_{j\ne k}^{ N} \lambda_a x_j$
are assumed to be known at the end user in the interactive decision-making process.
In a practical system the latter could be learned from the historical information on its edge computing delays.
In the simulations we assume that such information could be perfectly estimated at the end users.
It can be seen that the scheme with price equal to zero converges to
the NE point given by \emph{Theorem} \ref{theorem1}.
In contrast, the scheme with the optimal price converges to the SE point,
where the SE point is given by \emph{Theorem} \ref{theorem2}.
This accomplishes the goal that, via charging expenses for the edge computing service
the interactive decision-making process with self-interested users
converges to a Nash Equilibrium point equal to the Social Equilibrium point.
In this way, users' behavior is regulated effectively and the edge computing resources go to the users who value them the most.
Moreover, as compared with the latter scheme which needs around one hundred iterations to converge,
one can see that the pricing-based scheme converges very fast, i.e., with only several iterations.
The fast convergence property of the proposed pricing-based scheme
is especially useful when facing an online user-behavior learning problem.

\begin{figure}[!t]
\centering
\includegraphics[width=3in]{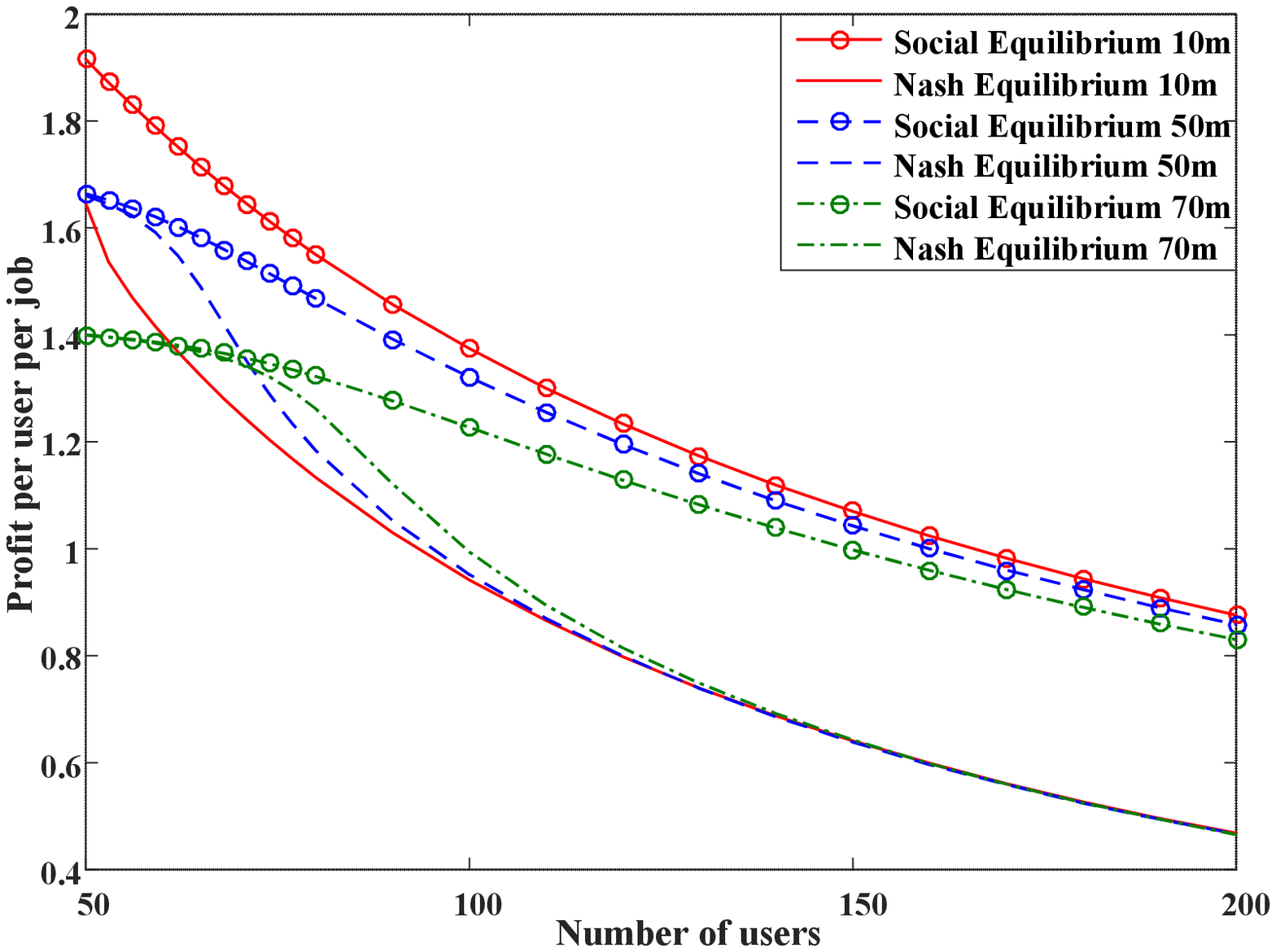}
 %where an .eps filename suffix will be assumed under latex,
% and a .pdf suffix will be assumed for pdflatex; or what has been declared via
\DeclareGraphicsExtensions. \caption{Profit per user per job at the equlibrium in the homogeneous user case.}
\label{fig4}
\vspace* {-6pt}
\end{figure}

\begin{figure}[!t]
\centering
\includegraphics[width=3in]{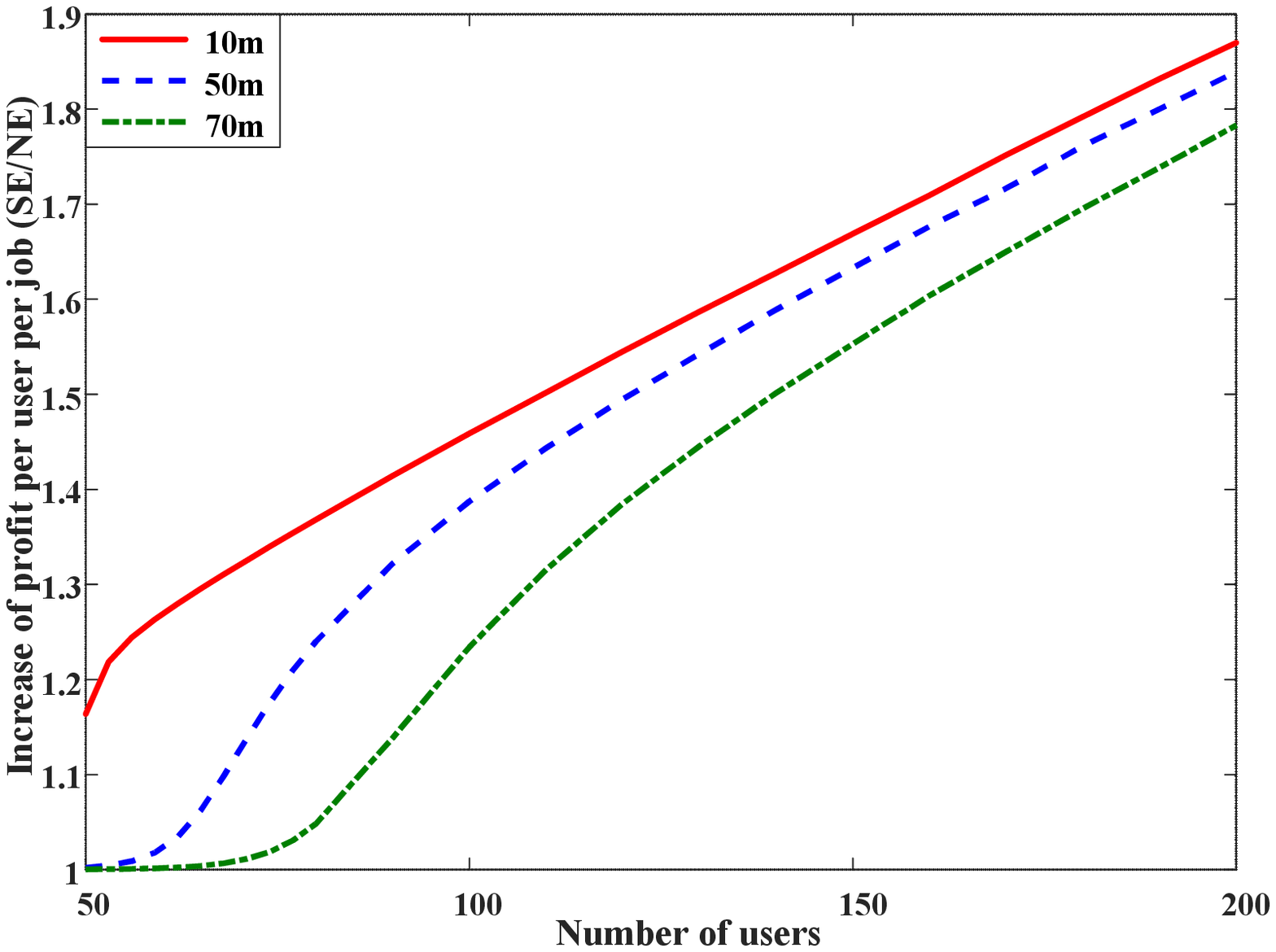}
 %where an .eps filename suffix will be assumed under latex,
% and a .pdf suffix will be assumed for pdflatex; or what has been declared via
\DeclareGraphicsExtensions. \caption{Profit Ratio SE/NE in the homogeneous user case.}
\label{fig5}
\vspace* {-6pt}
\end{figure}

Fig. \ref{fig4} illustrates the profit each user obtains by offloading a job to the
edge. As expected, at the SE point a greater profit is achieved, and this improvement
increases as the user moves closer to the AP.
This is because the congestion becomes more secious as the offloading increases, while pricing helps to combat that.
For the purpose of easy comparison, in Fig. \ref{fig5} we further plot the
ratio of the profit performance that we can get at the SE point and at the NE point,
which indicates how the network performance benefits when applying the optimal pricing-based scheme.
It can be seen that the ratio of SE/NE increases as the number of mobile device users connected to the AP increases.
For the case of 200 users and with the proposed optimal pricing-based scheme,
each user almost achieves a double profit as compared with applying the scheme charging no fees for
the mobile edge computing service.

\begin{figure}[!t]
\centering
\includegraphics[width=3in]{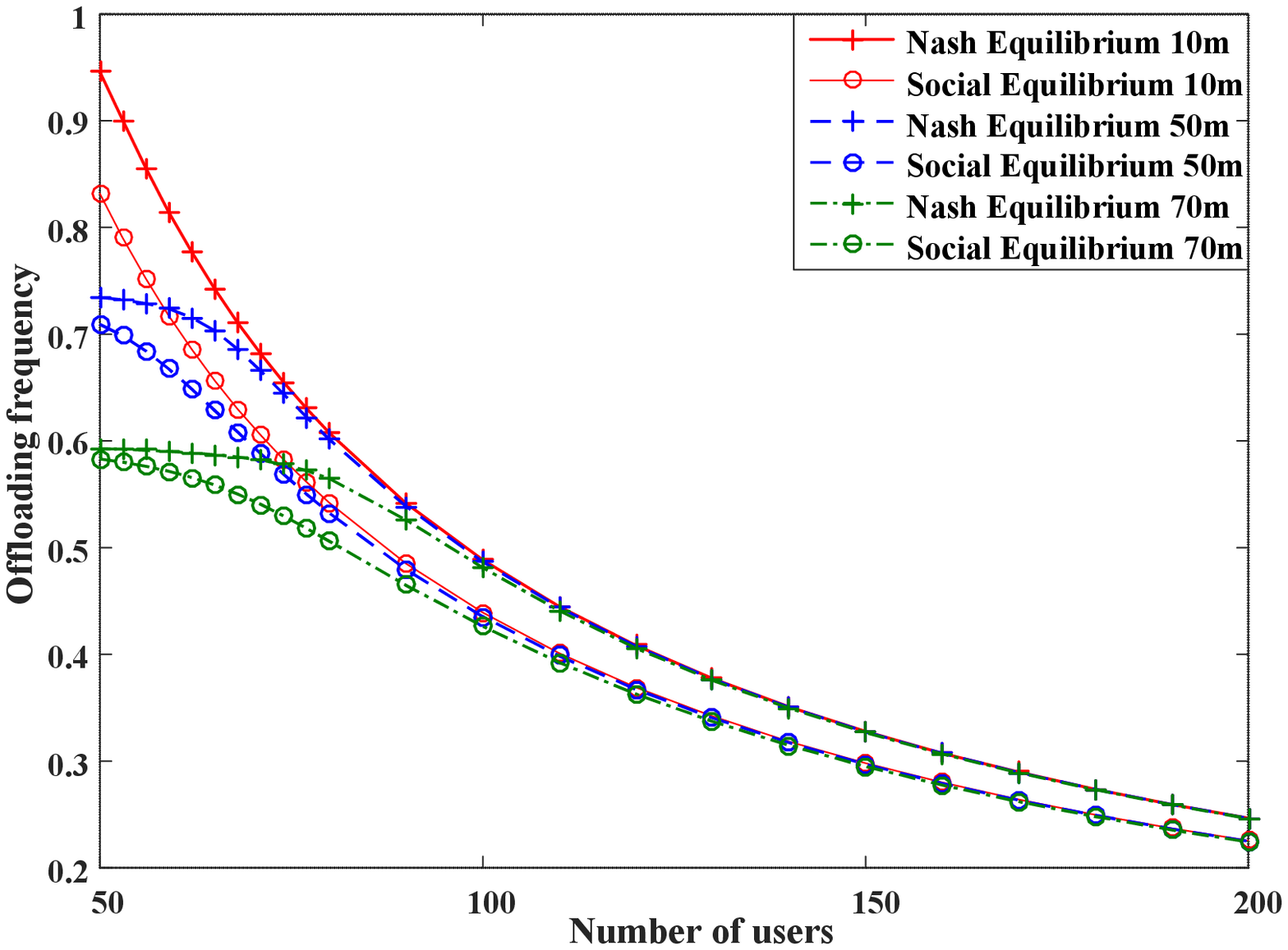}
 %where an .eps filename suffix will be assumed under latex,
% and a .pdf suffix will be assumed for pdflatex; or what has been declared via
\DeclareGraphicsExtensions. \caption{Offloading frequency at the equlibrium in the homogeneous user case.}
\label{fig6}
\vspace* {-6pt}
\end{figure}

In Fig. \ref{fig6} we plot the offloading frequency at the NE point and also at the SE point.
It can be seen that more jobs will be offloaded when an user moves closer to the AP,
since it experiences a smaller path loss and a better wireless channel.
On the other hand, less jobs will be offloaded at the SE point as compared with that at
the NE point. This is because, in the social-welfare game each user should pay for the congestion it causes to the other users.
This suggests that by charging some appropriate edge computing service fee, users' behaviors can be regulated effectively.
%As such, an user will keep its offloading amount under an appropriate amount for the other users' welfare.

\begin{figure}[!t]
\centering
\includegraphics[width=3in]{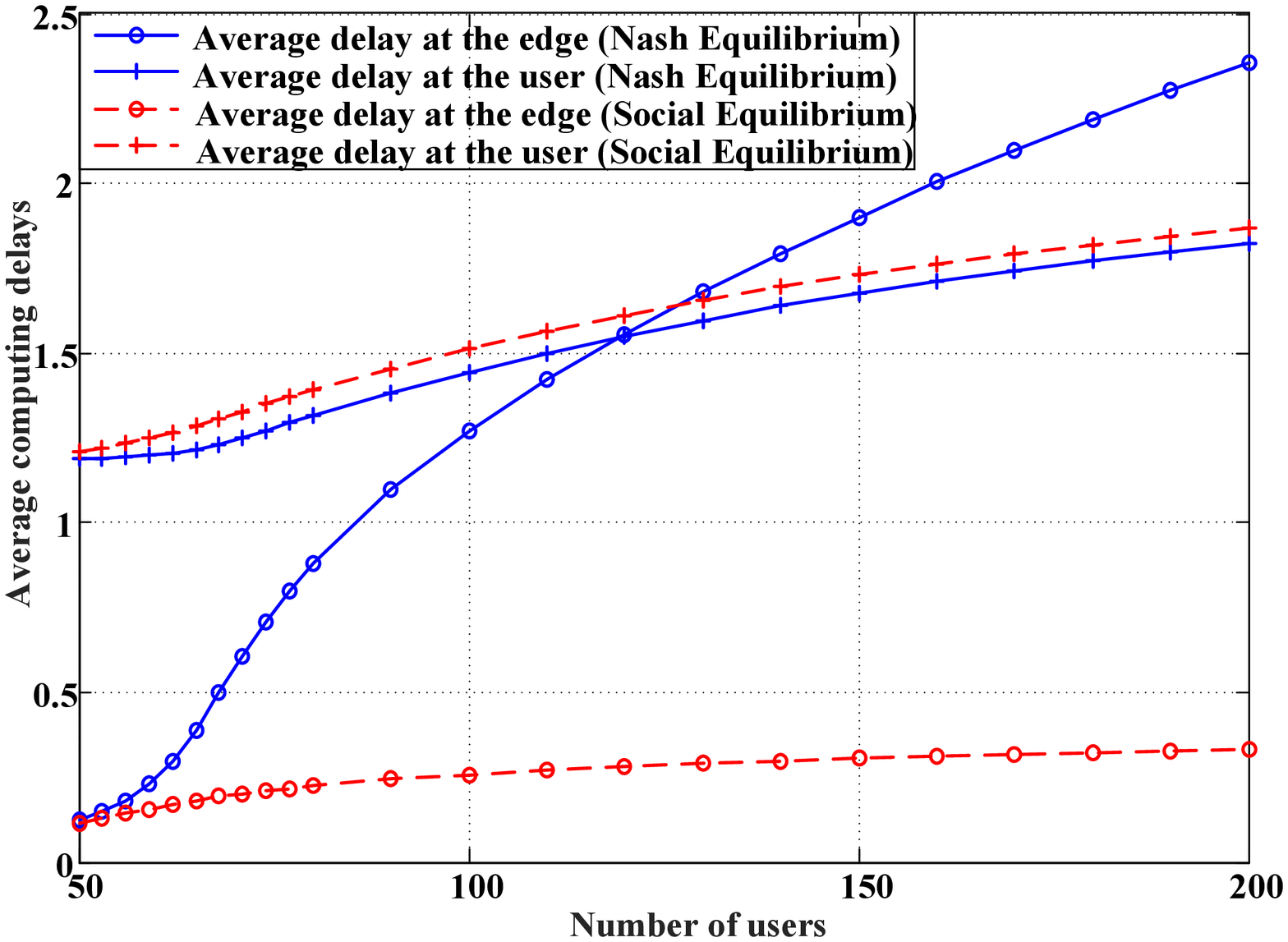}
 %where an .eps filename suffix will be assumed under latex,
% and a .pdf suffix will be assumed for pdflatex; or what has been declared via
\DeclareGraphicsExtensions. \caption{Average delays at the equlibrium in the homogeneous user case.}
\label{fig7}
\vspace* {-6pt}
\end{figure}

In Fig. \ref{fig7} we check the average delays of local computing as well as edge computing.
It can be seen that, the average delay of local computing increases a little bit
even if it takes into account the congestion it causes to the other users. In contrast,
the average delay of edge computing for the proposed pricing-based scheme
decreases a lot. %, as compared with applying the scheme charging no fees for the edge computing service
As such, at the SE point the user enjoys a smaller average delay.
In addition, this improvement increases as the number of mobile device users increases.
This should be expected, since in the social-welfare game the user offloads fewer jobs
to the edge, which reduces the burden of the edge in computing especially when there are a large number
of users.

\subsection{The general heterogeneous user case}
In the previous discussions, we have provided numerical results for the special case in which all the users
have the same utility functions.
In this subsection, we consider the heterogeneous user case where users are located at
a random different distance to the AP. Specifically,
we place mobile device users uniformly at random
on a ring of radius $10 \le d \le 75$ (unit: meters) and center located at the AP;
in this case and according to Fig. \ref{fig0}, one can see that the users have different utility functions.
%Moreover, the closer the user to the access point, the bigger its demand to offload is.
As such, it is difficult to derive closed-form results of the equilibrium as in the aforementioned homogeneous user case.
%Our goal in this subsection is to check if the interactive decision-making process among users still converges.

\begin{figure}[!t]
\centering
\includegraphics[width=3in]{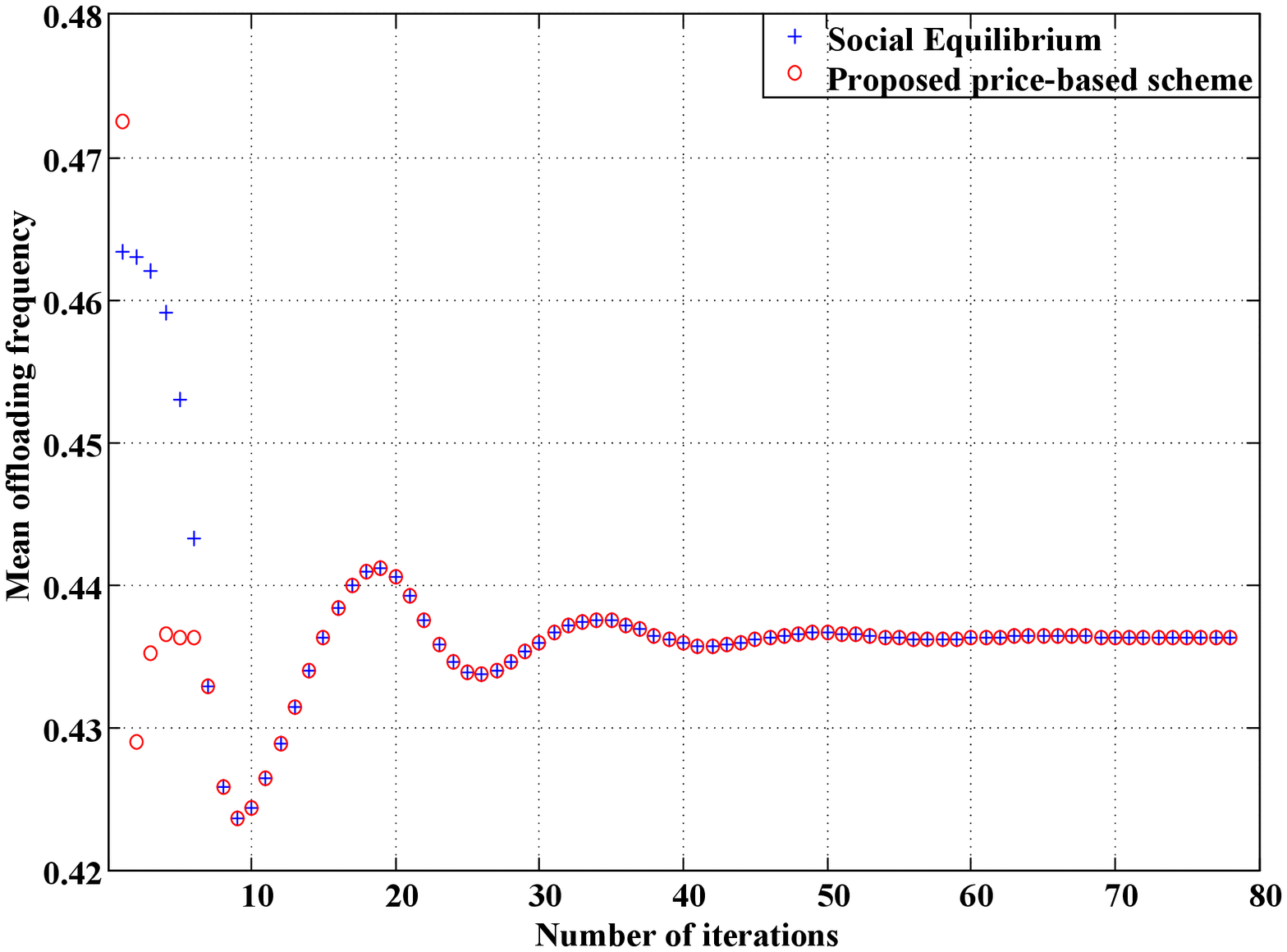}
 %where an .eps filename suffix will be assumed under latex,
% and a .pdf suffix will be assumed for pdflatex; or what has been declared via
\DeclareGraphicsExtensions. \caption{Convergence of the propsoed algorithm in the hetergeneous user case.}
\label{fig8}
\vspace* {-6pt}
\end{figure}

In Fig. \ref{fig8} we compare the convergence of the pricing-based scheme and that of the optimal Social Equilibrium.
The Social Equilibrium given by \emph{Theorem} \ref{theorem2} for the homogeneous user case can not be applied here since each user may
have a different utility function.
Instead, by some algebra derivations of (\ref{eq21}), we arrive at
\begin{align}
\mu_B-\sum\limits_{j\ne k}^{ N} \lambda_a x_j = F_S(x_k),\ k=1, \cdots, N,  \label{eq21b}
\end{align}
where $F_S(x_k)\triangleq\lambda_a x_k+\sqrt{{{c_0^t}\mu_B}/{ g_k (x_k)}}$.
The equations in (\ref{eq21b}) give the necessary conditions of the equilibrium of the game, indicating
the best response of a user when the other users' offloading decisions are given.
Based on (\ref{eq21b}) we update the offloading frequency $x_k$ one after another and iteratively;
by \emph{Definition 2}, if there exists an equilibrium point of this iteration, that
equilibrium should be the Social Equilibrium.
From Fig. \ref{fig8}, it can be seen that this iterative process converges after around
a hundred of rounds.
On the other hand, the optimal price given by \emph{Theorem} \ref{theorem3} also cannot be applied here. Instead, in the simulations we use the price of (\ref{eq27}), where
the offloading frequency of $\bar x_k^\star$ is given by the social problem.
It can be seen that the interactive decision-making process converges to the same
point as that by the social problem.

\begin{figure}[!t]
\centering
\includegraphics[width=3in]{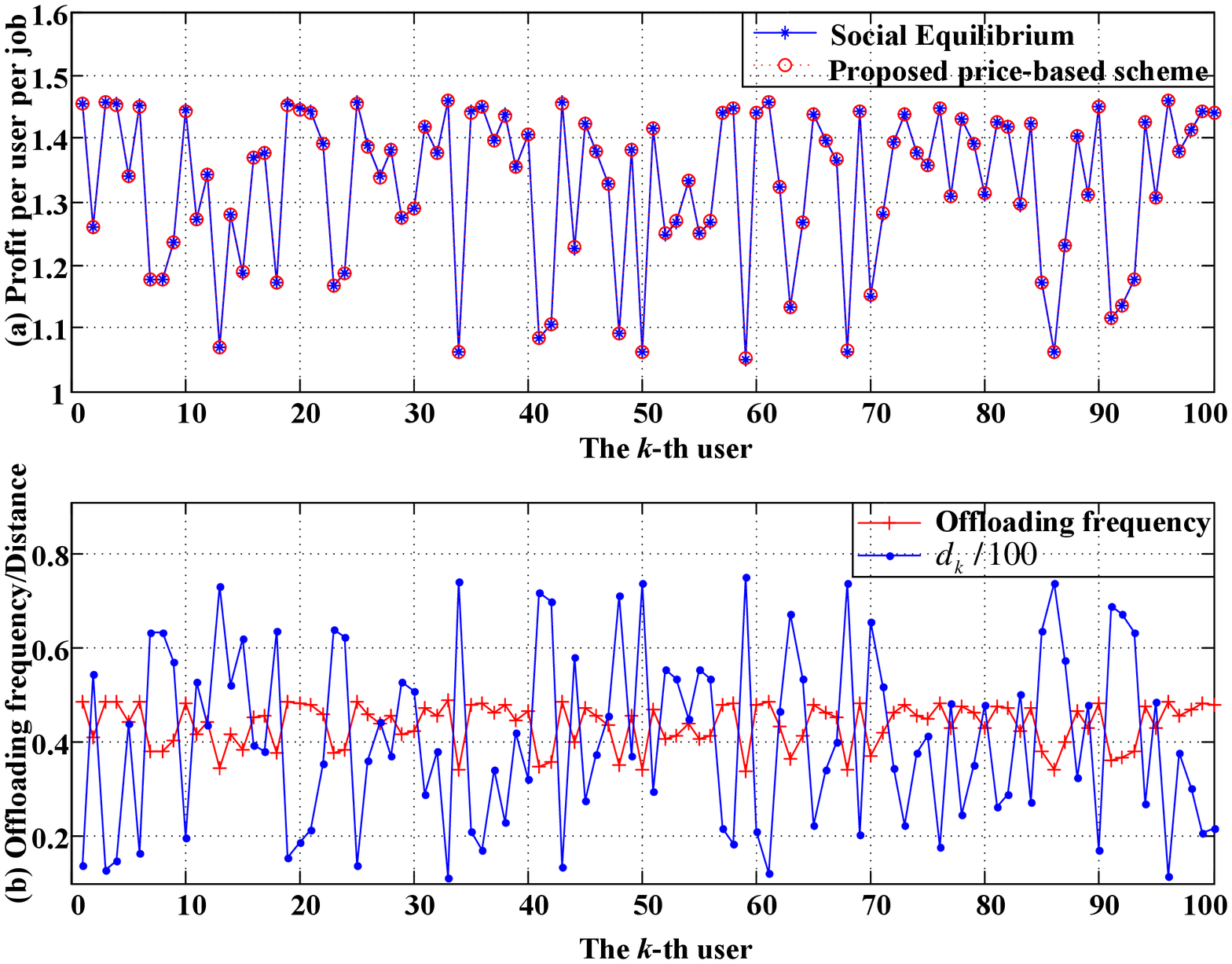}
 %where an .eps filename suffix will be assumed under latex,
% and a .pdf suffix will be assumed for pdflatex; or what has been declared via
\DeclareGraphicsExtensions. \caption{(a) Profit per user per job; (b) Offloading frequency at the equlibrium in the hetergeneous user case.}
\label{fig9}
\vspace* {-6pt}
\end{figure}

Fig. \ref{fig8} plots the mean offloading frequency of users.
In Fig. \ref{fig9} we further respectively plot the profit each user obtains and its offloading
frequency at the equilibrium. Interestingly, it can be seen that by the proposed pricing-based scheme and at the equilibrium, each user
obtains the same profit as that at the Social Equilibrium. According to \emph{Theorem} \ref{theoremA},
that Social Equilibrium should be unique. Hence, the proposed pricing-based scheme is global optimal,
under which the interactive decision-making process among users converges
to a NE point equal to the SE point. In addition, as expected, the user closer to the AP offloads more frequently.

\section{Conclusion}
In this paper, we have proposed an incentive-based offloading control framework for the mobile edge computing network,
which consists of a radio access network (RAN) that is equipped with a mobile edge server (MEC) of finite computing power, and
serves multiple resource-hungry mobile device users by charging
users appropriate economic expenses for offloading computation jobs to the edge.
Firstly, we have introduced a novel physical layer rooted utility function, which measures
the reduction in cost by offloading as compared to executing jobs locally.
We have proved this utility function satisfies some basic economic properties, thus
bridging the gap between physical layer optimizations and economics.
With this proposed utility function we then formulate two computation offloading decision making games,
with one maximizing the individual's interest and the other maximizing the overall system's welfare.
We have addressed analytically the structural property of the games and admit in closed form
the Nash Equilibrium and the Social Equilibrium, respectively, based on which we then
proposed an optimal pricing scheme, which regulates users' action by charging expenses for providing the edge computing service.
Numerical results support our theoretical results that under our proposed optimal pricing-based scheme,
the interactive decision-making process with self-interested users
converges to a Nash Equilibrium point equal to the Social Equilibrium point.

\appendices
\section{Proof of \emph{Lemma 1}} \label{appA}
By definition, the demand function is equal to the derivative of the utility function with respect to the offloading frequency $x_k$, which, combined with (\ref{eq13})
indicates that
\begin{align}
g_k({x}_{k})=c_k^e  E_k^{\rm LC} -  \dfrac{\eta}{\beta_k({x_k})}+\dfrac{\eta\beta_k^\prime({x_k})x_k}{\beta_k^2(x_k)}+ \dfrac{c_k^t \mu_m}{(\mu_m-\lambda_a\bar{x}_k)^2} \label{eqA1}
\end{align}
where $\eta \triangleq (c_k^t+c_k^e P_t) l_a\mu_a$.

Looking into the expressions in (\ref{eqA1}), the second term $-  \dfrac{\eta}{\beta_k({x_k})}$ as well as the last term $ \dfrac{c_k^t \mu_m}{(\mu_m-\lambda_a\bar{x}_k)^2}$
decreases as $x_k$ increases, since $\beta_k(x_k)$ is a decreasing function with respect to $x_k$ (see (\ref{eq3})).
In addition, it holds that $-x\beta^\prime(x)>0$. And in the sequel, we will further show that $-x_k\beta_k^\prime({x_k})$ increases
as $x_k$ increases, thus giving the proof that $g_k({x}_{k})$ decreases as $x_k$ increases.

Letting $\psi(x)=-x\beta^\prime(x)$ and by the derivative of $\psi(x)$, we arrive at
\begin{align}
\psi^\prime(x)&=\lim_{\Delta x\rightarrow 0} \dfrac{\psi(x+\Delta x)-\psi(x)}{\Delta x} \nonumber\\
&=\lim_{\Delta x\rightarrow 0}- \dfrac{(x+\Delta x)\beta^\prime(x+\Delta x)-x \beta^\prime(x)}{\Delta x}.  \label{eqA2}
\end{align}
Besides this, by the definition of derivative it holds that
\begin{subequations}
\begin{align}
\beta^\prime(x) &=\lim_{\Delta x\rightarrow 0} \dfrac{\beta(x+\Delta x)-\beta(x)}{\Delta x},  \label{eqA3a} \\
\beta^\prime(x+\Delta x) &=\lim_{\Delta x\rightarrow 0} \dfrac{\beta(x+\Delta x-\Delta x)-\beta(x+\Delta x)}{-\Delta x}.   \label{eqA3b}
\end{align}
\end{subequations}
Combining (\ref{eqA3a}) and (\ref{eqA3b}) yields
\begin{align}
(x+\Delta x)\beta^\prime(x+\Delta x)-x \beta^\prime(x)= \beta(x+\Delta x)-\beta(x).  \nonumber
\end{align}
This, combined with (\ref{eqA2}), indicates that
\begin{align}
\psi^\prime(x)&= -\beta^\prime(x).  \nonumber
\end{align}
Since $\beta_k(x_k)$ is a decreasing function with respect to $x_k$, it holds true that $\psi^\prime(x)>0$.
Hence, $-x\beta^\prime(x)$ is a monotonically increasing function with respect to $x$.
This completes the proof of the first part of \textit{Lemma 1}.

According to the above argument, it shows that $g_k(x_k)$ is a monotonically decreasing function with respect to $x_k$.
In addition, it can be verified that
\begin{align}
\lim_{x_k \rightarrow 0^+}g_k(x_k)  &= c_k^e  E_k^{\rm LC} + \dfrac{c_k^t \mu_m}{(\mu_m-\lambda_a)^2}>0,  \nonumber\\
\lim_{x_k \rightarrow 1^-}g_k(x_k)  &= - \infty.  \nonumber
\end{align}
Therefore, the solution of the equation $g_k(x_k)=0$ exists and it is unique.
This completes the proof.

\section{Proof of \emph{Theorem 1}} \label{appB}
By definition, at the equilibrium the derivative of the objective function should be zero.
Hence, according to (\ref{eq18}) and at the equilibrium it holds that
\begin{align}
(\mu_B-\sum\nolimits_{j=1}^{ N} \lambda_a x_j)^2= \frac{c_0^t (\mu_B-\sum\nolimits_{j \ne k}^{ N} \lambda_a x_j)}{ g_0(x_k)},\ \forall k.  \label{eqB1}
\end{align}

In what follows, we prove by contradiction and show that for any given offloading frequency vector ${\bf x}=(x_1, \cdots, x_N)$ satisfying (\ref{eqB1}),
$x_m=x_n$ holds true for $\forall m, n$.
Substituting this result into (\ref{eqB1}), we arrive at the result of (\ref{eq19}) in Theorem \ref{theorem1}.

Assume that there exists a pair of $(x_m, x_n)$, with $x_m > x_n$.
By \textit{Lemma} \ref{lemma1}, $g_0(x_k)$ is a monotonically decreasing function.
As such, it holds that $g_0(x_m)<g_0(x_n)$.
Besides, it holds that $\mu_B-\sum\nolimits_{j \ne m}^{ N} \lambda_a x_j>\mu_B-\sum\nolimits_{j\ne n}^{ N} \lambda_a x_j$,
since $(\mu_B-\sum\nolimits_{j=1}^{ N} \lambda_a x_j)+\lambda_a x_m>(\mu_B-\sum\nolimits_{j=1}^{ N} \lambda_a x_j)+\lambda_a x_n$.
Therefore, we have
\begin{align}
\frac{c_0^t (\mu_B-\sum\nolimits_{j \ne m}^{ N} \lambda_a x_j)}{ g_0(x_m)}>\frac{c_0^t (\mu_B-\sum\nolimits_{j \ne n}^{ N} \lambda_a x_j)}{ g_0(x_n)}. \nonumber
\end{align}
This contradicts with (\ref{eqB1}).
Similarly, we can prove that the assumption $x_m<x_n$ also contradicts with (\ref{eqB1}).
In summary, it holds that $x_m=x_n$, $\forall m, n$.
This completes the proof.

\section{Proof of \emph{Corollary 1}} \label{appC}
For the ease of exposition, let the left side of Eq. (\ref{eq19}) denote by
\begin{align}
\phi(x_k)= N\lambda_a x_k +\sqrt{\dfrac{c_0^t(\mu_B-(N-1)\lambda_a x_k)}{g_0(x_k)}}. \nonumber
\end{align}
It can be verified that $\phi(x_k)$ is a monotonically increasing function with respect to $x_k$.
In addition, we will show later that there exists some $x_k$ such that
$\phi(x_k)$ is greater than the right hand side of Eq. (\ref{eq19}), i.e., $\mu_B$.
As such, as long as there exists some $x_k$ such that $\phi(x_k)$ is less than $\mu_B$,
we can infer that there exists a unique solution to Eq. (\ref{eq19}).

Specifically, it can be verified that
\begin{align}
&\lim_{x_k \rightarrow x_k^{up}} \phi(x_k)= + \infty > \mu_B, \lim_{x_k \rightarrow 0} \phi(x_k)= \sqrt{\dfrac{c_0^t\mu_B}{g_0(0)}}.\nonumber
\end{align}
Therefore, we arrive at a sufficient condition for the existence of solutions to Eq. (\ref{eq19}) as follows,
\begin{align}
\sqrt{\dfrac{c_0^t\mu_B}{g_0(0)}} < \mu_B \Leftrightarrow \dfrac{c_0^t}{\mu_B} < g_0(0)=c_0^e  E_k^{\rm LC} + \dfrac{c_0^t \mu_m}{(\mu_m-\lambda_a)^2}. \nonumber
\end{align}
And, that solution should be unique due to the monotonicity of $\phi(x_k)$ with respect to $x_k$.
This completes the proof that if (\ref{eq20}) holds true, there exists a positive unique Nash Equilibrium.

In the sequel, we prove that if $g_0(0) \le {c_0^t}/{\mu_B} $, the profit function $V_k({\bf x})$
is a non-increasing function with respect to $x_k$. In that case, the best
response for the mobile device user is to stop offloading and compute all the jobs locally.

Some observations are in order. Firstly, it can be verified that
\begin{align}
\frac{c_0^t (\mu_B-\sum\nolimits_{j \ne k}^{ N} \lambda_a x_j)}{(\mu_B-\sum\nolimits_{j=1}^{ N} \lambda_a x_j)^2} \ge
\frac{c_0^t }{(\mu_B-\sum\nolimits_{j=1}^{ N} \lambda_a x_j)} \ge
\dfrac{c_0^t}{\mu_B}. \nonumber
\end{align}
Secondly, by assumption it holds that $g_0(0) \le \dfrac{c_0^t}{\mu_B} $.
Combining the above observations yields
\begin{align}
\frac{c_0^t (\mu_B-\sum\nolimits_{j \ne k}^{ N} \lambda_a x_j)}{(\mu_B-\sum\nolimits_{j=1}^{ N} \lambda_a x_j)^2} \ge
g_0(0) \ge g_0(x_k). \nonumber
\end{align}
This, combined with (\ref{eq12b})(\ref{eq18}), indicates that the derivative of the profit function $V_k({\bf x})$
is no more than zero. This completes the proof.

\section{Proof of \emph{Theorem 2}} \label{appD}
As discussed in Sec. IV. B, the equations in (\ref{eq21}) give the necessary conditions of the equilibrium of the game.
Hence, for the purpose of studying the existence and the uniqueness of the Social Equilibrium point,
it is sufficient to investigate the solution of the equations in (\ref{eq21}).

The combination of the $N$ equations in (\ref{eq21}) indicates that at the equilibrium there exists some positive value of $t$ such that
$g_k (x_k)=t, \forall k$.
As such, the equations in (\ref{eq21}) can be reformulated as follows,
\begin{subequations}
\begin{align}
&\mu_B-\sum\nolimits_{j=1}^{ N} \lambda_a x_j = \sqrt{\dfrac{{c_0^t}\mu_B}{ t}}, \label{eqD1a}\\
&g_k (x_k)=t, \forall k. \label{eqD1b}
\end{align}
\end{subequations}

In the sequel, we will first prove the existence of the solutions of (\ref{eqD1a})(\ref{eqD1b}), followed by the
proof of its uniqueness. In this way, we give the proof of \emph{Theorem} \ref{theoremA}

\emph{There exists at least a solution to the equations in (\ref{eqD1a})(\ref{eqD1b}):}
Some observations are in order.
\begin{enumerate}
\item Firstly, the nontrivial case where users cannot benefit from offloading is out of the scope of this paper.
%According to \emph{Corollary} \ref{corollary1} and \emph{Corollary} \ref{corollary2},
As such, only if a user's demand at $x_k=0$ is greater than the edge computing cost, will it choose to offload its
job to the edge i.e., it holds true that
\begin{align}
g_k(0)> \frac{c_0^t }{\mu_B}
\Leftrightarrow \mu_B>  \sqrt{\dfrac{{c_0^t}\mu_B}{g_k (0)}}. \nonumber
\end{align}
Therefore, at the offloading vector of ${\bf x}={\bf 0}$ it holds that
\begin{align}
\mu_B-\sum\nolimits_{j=1}^{ N} \lambda_a x_j >  \sqrt{\dfrac{{c_0^t}\mu_B}{t}}. \label{eqD0}
\end{align}
%In what follows, we prove the inequalities in (\ref{eqD0}) by contradiction. Assume that $\exists i$ at which
%\begin{align}
%&\mu_B-\sum\nolimits_{j=1}^{ N} \lambda_a x_j < \sqrt{\dfrac{{c_0^t}\mu_B}{ g_i (x_i) }} \nonumber \\
%\Leftrightarrow & g_i (x_i) < \dfrac{{c_0^t}\mu_B}{(\mu_B-\sum\nolimits_{j=1}^{ N} \lambda_a x_j)^2 }. \nonumber
%\end{align}
%Since $ g_i (x_i)$ is a monotonically decreasing function and the right hand of the above inequality is a monotonically
%increasing function of $x_i$, we could always find another $x_i^\prime < x_i$ so as to
%\begin{align}
%g_i (x_i) <  g_i (x_i^\prime)&= \dfrac{{c_0^t}\mu_B}{(\mu_B-\sum\nolimits_{j\ne i}^{ N} \lambda_a x_j-\lambda_a x_i^\prime)^2 } \nonumber\\
%&< \dfrac{{c_0^t}\mu_B}{(\mu_B-\sum\nolimits_{j=1}^{ N} \lambda_a x_j)^2 }, \nonumber
%\end{align}
%which indicates that all the users benefit (achieves a bigger profit)
%from an unilaterally change of the $i$-th user's offloading strategy to $x_i^\prime$.
%This contracts with the definition of Social Equilibrium. This completes the
%proof of Eq. (\ref{eqD0}).
\item Secondly, it holds that $\lim_{t \rightarrow 0^+}  \sqrt{\dfrac{{c_0^t}\mu_B}{t}}=  +\infty $.
As such, there exists some offloading vector of $\bf x$ so as to
\begin{align}
\mu_B-\sum\nolimits_{j=1}^{ N} \lambda_a x_j <  \sqrt{\dfrac{{c_0^t}\mu_B}{t}}. \label{eqD01}
\end{align}
\item Thirdly, the left hand side of Eq. (\ref{eqD1a}) is monotonically decreasing while the right hand side is monotonically increasing.
\end{enumerate}
Combining the above observations, we conclude that there exists at least a solution
to the equations in (\ref{eqD1a})(\ref{eqD1b}).

\emph{The solution of the equations in (\ref{eqD1a})(\ref{eqD1b}) is unique:} We give the proof by contradiction.
Assume that there exist two equilibrium points ${\bf x}^1=(x_1^1, \cdots, x_N^1)$ and ${\bf x}^2=(x_1^2, \cdots, x_N^2)$, at which
it satisfies that $t_1=g_k(x_k^1)$, $\forall k$, $t_2=g_k(x_k^2)$, $\forall k$, and $t_1 \ne t_2$.
Without loss of generality, let's consider the case of $t_1 > t_2$. As such, we have $g_k(x_k^1)>g_k(x_k^2)$, $\forall k$.
In addition, by \emph{Lemma} \ref{lemma1} the demand function $g_k({x}_{k})$ is a monotonically
decreasing function of the offloading frequency $x_k$.
Therefore, it holds that $x_k^1 < x_k^2$, $\forall k$, which further indicates that
\begin{align}
&\mu_B-\sum\nolimits_{j=1}^{ N} \lambda_a x_j^1 > \mu_B-\sum\nolimits_{j=1}^{ N} \lambda_a x_j^2, \nonumber \\
\Leftrightarrow  &\sqrt{\dfrac{{c_0^t}\mu_B}{ t_1}} > \sqrt{\dfrac{{c_0^t}\mu_B}{ t_2}}
\Leftrightarrow  t_1 < t_2. \label{eqD2}
\end{align}
The inequality in (\ref{eqD2}) contradicts with the assumption of $t_1 > t_2$.
This completes the proof.

\section{Proof of \emph{Theorem 4}} \label{appE}
Substituting (\ref{eqP1}) into \emph{Theorem} \ref{theorem1} indicates that at the Nash Equilibrium it holds that
\begin{align}
N\lambda_a \hat x_0^\star +\sqrt{\dfrac{c_0^t(\mu_B-(N-1)\lambda_a \hat x_0^\star)}{g_0(\hat x_0^\star)-P}}=\mu_B.  \label{eqE2}
\end{align}
For the purpose of having a Nash Equilibrium point equal to the Social Equilibrium point, we should have $\hat x_0^\star=\bar x_0^\star$.
As such, the solution $\hat x_0^\star$ should also satisfy Eq. (\ref{eq22}).
Combining Eq. (\ref{eqE2}) and Eq. (\ref{eq22}) yields
\begin{align}
& {\dfrac{c_0^t(\mu_B-(N-1)\lambda_a \bar x_0^\star)}{g_0(\bar x_0^\star)-P}}= {\dfrac{c_0^t \mu_B}{g_0(\bar x_0^\star)}} \nonumber\\
\Leftrightarrow& P= \dfrac{(N-1)\lambda_a\bar x_0^\star}{\mu_B}{g_0(\bar x_0^\star)}. \nonumber
\end{align}
This completes the proof.

\bibliography{mybib}
\bibliographystyle{IEEEtran}

\end{document}